%% file: hydro-fin.tex
\documentclass{elsart}
\usepackage{amsfonts}\usepackage{amsbsy}
\usepackage{feynmp}
\usepackage{graphicx}
\def\cal{\mathcal}
\input{entr-mac.tex}

\begin{document}
\begin{frontmatter}
\title{\vspace*{-15mm} Hydrodynamics   of Resonances }

\author{D. N. Voskresensky$^{1,2}$}

\maketitle

\noindent $^{1}${\it Gesellschaft f\"ur Schwerionenforschung mbH,
Planckstr. 1, 64291 Darmstadt, Germany}
\\
$^{3}${\it National Research Nuclear University "MEPhI",
Kashirskoe sh. 31, Moscow 115409, Russia}

\begin{abstract}
We derive  system of equations  describing
 fluidity  of the medium consisting of non-relativistic particles with finite
mass-widths. For that we use expressions for the kinetic Noether
4-current and the Noether energy-momentum tensor being conserved
provided  one uses self-consistent approximations to the gradient
expanded Kadanoff-Baym equations. Kinetic coefficients entering
equations of non-ideal hydrodynamics of resonances are obtained
 in terms of the
 real and imaginary parts of the self-energies within a relaxation time approximation.
\end{abstract}
\end{frontmatter}
\begin{fmffile}{cons-fig}

\section{Introduction}\label{Intr}

The appropriate frame for the description of non-equilibrium
processes is the real-time formalism of quantum field theory, see
\cite{Schw,KB62,B62,Keld64,D84,BM90,VBRS95,KV96} and refs.
therein. This formalism finds now applications in many fields. The
reason is the necessity of the dynamical description of broad
resonances, as well as stable particles, which acquire a
considerable width because of collision broadening. E.g., off-mass
shell particles and resonances are extensively produced in heavy
ion collisions.

Description of particles with broad widths requires development of
self-consistent schemes with the conservation laws, being at least
approximately satisfied
\cite{B62,IKV99,IKV00,IKHV01,KIV01,KRIV06,IV09}. Refs.
\cite{KIV01} have shown that for the generalized kinetic equation
in the so called Kadanoff-Baym (KB) form derived within the first
space-time gradient approximation the conservation laws are
exactly satisfied, provided one uses the  $\Phi$-derivable
approximations, whereas in the so called Bottermans-Malfliet (BM)
form they are approximately fulfilled within consistent first
order gradient expansion. Such approaches permitted a numerical
transport treatment of the off-shell dynamics of the particles in
the matter \cite{EM99,CJ00,L00,IV09}.

Very close to the equilibrium at times much larger than the
typical time for changes of kinetic quantities the kinetic
description can be replaced by a more economical  hydrodynamical
 description. The fluid-dynamical approach is fairly efficient for
description of heavy-ion collisions in a broad collision energy
range from SIS to RHIC energies (see e.g. \cite{SG,IR,IdealRHIC}).
Recently an interest in the transport coefficient issue  has been
sharply  increased in  heavy-ion collision physics. Large values
of the elliptic flow $v_2$ were observed at very high collision
energies, at RHIC~\cite{RHIC-v2}. This can be understood as the
created QGP behaves as a fluid with a  small (but non-zero) value
of the shear viscosity-to-entropy density ratio. The statement was
confirmed with the help of non-ideal 2-dimensional hydrodynamical
simulations, see~\cite{IdealRHIC}.  In order to describe data in a
broad energy range from SIS to SPS  the 2-dimentional calculations
are insufficient. The existing 3-dimensional hydrodynamical
schemes use up to now ideal hydrodynamics. Although viscosity
effects prove to be large, they are simulated indirectly with the
help of an artificially introduced friction between different
components of the liquid, see \cite{IR}.  Refs.
\cite{SV09,R09,SV10,R10} argue that viscosity and heat
conductivity effects are very important
 in the dynamical description  of the first-order phase
transitions. Search of possible manifestations of the critical
endpoint in the nuclear matter phase diagram is one of the
intriguing perspectives of the projects at FAIR, NICA and at low
energy RHIC campaign. All mentioned stimulates significant
interest to the development of the generalized fluid-dynamical
description of the resonance matter.

In the modeling of the strongly interacting matter,  interactions
are often treated within the quasiparticle approximation although
the width effects at least for some particle species can be very
large. Refs.~\cite{SR08,SR09,KTV09} calculated the shear and bulk
viscosities of the hadron and quark phases within the
quasiparticle approach in the relaxation time approximation in
case where the effective masses of the constituents depend on the
temperature and on the baryon density.

Although it is well known that the generalized kinetic approach
naturally leads to expressions for the transport coefficients in
terms of correlators like those discussed by Kubo, see
\cite{Martin}, subsequent calculation of  these  correlators
presents a complicated problem. To the best of our knowledge there
are no consistent derivations of the kinetic coefficients from the
generalized kinetic scheme which would be presented in terms of
real and imaginary parts of the particle Green functions and
self-energies, beyond the scope of the quasiparticle
approximation. Estimates of the width effects on viscosities based
on some reasonable conjectures have been done only recently
\cite{BS06,KTV09}.

In this paper using expressions for the kinetic Noether 4-current
and energy-momentum tensor, which we introduce in sect. 2
following \cite{IKV00}, we derive the generalized fluid-dynamical
equations for the description of  resonances (see sect. 3). In
integral form these equations are presented in Appendix A. To be
specific starting from sect. 3 we focus on description of
non-relativistic dynamics. Then in sect. 4 we find an approximate
solution  of the kinetic equation in  the BM form and in sect. 5
we derive transport coefficients expressed in terms of the
self-energy functions. Some details necessary for calculations of
kinetic coefficients are deferred to Appendices B-F. Thus we
construct a consistent hydrodynamical approach for the description
of particles with mass-widths.

\section{Preliminaries}

\subsection{Generalized kinetic quantities in physical notations}
\label{Notations}

To avoid the imaginary factors inherent in the standard Green
function formulation one can introduce quantities, which are real
and, in the quasi-homogeneous limit, positive, having clear
physical meaning. So, instead of Green functions $\Gr^{ij}(x,p)$
and self-energies $\Se^{ij}(x,p)$ ($i,j\in\{-+\}$) in the Wigner
representation, we use the kinetic notation of   Ref.
\cite{IKV00}, namely
%
\begin{eqnarray}
\label{F}
\Fd (x,p) &=& \A (x,p) \fd (x,p)
 =  (\mp )\ii \Gr^{-+} (x,p) , \nonumber\\
\Fdt (x,p) &=& \A (x,p) [1 \mp \fd (x,p)] = \ii \Gr^{+-} (x,p) ,
\end{eqnarray}
%
for the generalized Wigner functions $\F$ and $\Ft$ and the
corresponding 4-phase-space distribution functions $\fd(x,p)$ and
Fermi/Bose factors $[1 \mp \fd (x,p)]$. The upper sign corresponds
to fermions, while the lower sign, to bosons. The space-time
variables are $x\equiv x^{\mu}=(t ,\vec{r}),$  $t =\frac{1}{2}(t_1
+t_2 )$, $\vec{r}=\frac{1}{2}(\vec{r}_1 +\vec{r}_2 )$, and the
Fourier transformed $\xi =x_1 -x_2$ variables are $p\equiv
p^{\mu}$.

 The spectral function
(spectral density) is
%
\begin{eqnarray}
\label{A}
 A (x,p) \equiv -2\Im \Gr^R (x,p) = \Fdt \pm \Fd ,
\end{eqnarray}
and $\Gr^R$ is the retarded propagator. The spectral function
satisfies the sum-rule
\begin{eqnarray} \label{sumA}
 \int_{-\infty}^{\infty}  \frac{\di p_0}{2\pi} A
(x,p)=1 ,
\end{eqnarray}
for non-relativistic particles or
\begin{eqnarray} \label{sumAr}
 \int_{-\infty}^{\infty}  \frac{\di p_0}{2\pi} p_0 A
(x,p)=1 ,
\end{eqnarray}
for relativistic bosons.

The reduced gain and loss rates of the collision integral are
defined as
%
\begin{eqnarray}
\label{gain} \Ldt (x,p) =   (\mp )\ii \Se^{-+} (x,p),\quad \Ld
(x,p)  =  \ii \Se^{+-} (x,p),
\end{eqnarray}
%
with the damping width
%
\begin{eqnarray}
\label{G-def}
\Gamma (x,p)&\equiv& -2\Im \Se^R (x,p) = \Ld (x,p)\pm\Ldt (x,p),
\end{eqnarray}
%
where $\Se^R$ is the retarded self-energy.

\subsection{Kadanoff--Baym and Botermans-Malfliet forms of generalized kinetic equation}\label{KB}

 For simplicity consider the case, when there are
no external fields. In terms of the generalized particle
distribution function $F(x,p)$ the KB equation requires the form
(so called KB form),
%
\begin{eqnarray}
\label{keqk1}
\Do
\Fd (x,p) -
\Pbr{\Ldt , \Re\Gr^R} &=& C (x,p) .
\end{eqnarray}
%
Here the differential drift operator is
%
\begin{eqnarray}\label{Drift-O}
\Do =\left(\vu_{\mu} -
\frac{\partial \Re\Sa^R}{\partial p^{\mu}}
\right)\partial^{\mu}_x +
\frac{\partial \Re\Sa^R}{\partial x^{\mu}} \frac{\partial
}{\partial p_{\mu}}
\quad\mbox{with}\quad v^\mu=\frac{\partial}{\partial
p_{\mu}}G_0^{-1}(p),
\end{eqnarray}
%
where $\Gr^{-1}_{0}(p)$ is the Fourier transform of the inverse
free Green function
\begin{eqnarray}
\label{G0} \Gr^{-1}_{0}(p)=\left\{
\begin{array}{ll}
p^2-m^2\quad&\mbox{for relativistic bosons}\\ p_0-m-{\vec
p}^2/(2m)\quad&\mbox{for non-rel. fermions or bosons.}
\end{array}\right.
\end{eqnarray}
In non-relativistic case we  count the energy and the chemical
potential from the mass. For relativistic bosons $v^\mu =2p^{\mu}$
and for non-relativistic particles of the mass $m$,
\begin{eqnarray}\label{n-relvel}
v^{\mu}\simeq (1,\vec{p}/m ).
\end{eqnarray}
Symbol $\Pbr{...,...}$ denotes the standard Poisson bracket,
\begin{eqnarray}
\label{[]}
\Pbr{f(x,p) , \varphi(x,p)} =
\frac{\partial f}{\partial p^{\mu}}
\frac{\partial \varphi}{\partial x_{\mu}}
-
\frac{\partial f}{\partial x^{\mu}}
\frac{\partial \varphi}{\partial p_{\mu}},
\end{eqnarray}
%
in covariant notation. Acting on an arbitrary function $\Psi$ the
drift operator $\Do$ yields
\begin{eqnarray}\label{DoM}
\Do\Psi =\Pbr{M,\Psi},
\end{eqnarray}
with the  ``mass'' function
%
\begin{eqnarray}\label{M}
M(x,p)=\Gr^{-1}_{0}(p) -\Re\Se^R (x,p).
\end{eqnarray}
 The drift $\Do$-term describes the particle drag
flow. The commutator term in (\ref{keqk1}) has no clear physical
meaning partially relating to the back flow and fluctuation
effects. The collision term
%
\begin{eqnarray}
\label{Coll(kin)}
C (x,p) =
\Ldt (x,p) \Ft (x,p)
- \Ld (x,p) \F (x,p)
\end{eqnarray}
%
has the local part, $C_{\rm loc}(x,p)$, and also the memory
correction $C_{\rm mem}(x,p)$. The latter  appears if one includes
into consideration the self-energy diagrams with more than two
vertices, cf. \cite{IKV00}. Note that in the local approximation
the collision term is charge (e.g., the baryonic number) and
energy--momentum conserving by itself \footnote{Here and below the
$\Tr$ means a sum over all possible internal degrees of freedom,
like spin, and over possible particle species.}
%
\begin{eqnarray}
\label{C-conser}
\mbox{Tr} \int \dpi{p}
\left(
\begin{array}{lll}
e\\p^\mu
\end{array}
\right) C^{\scr{loc}}=0.
\end{eqnarray}
The kinetic equation (\ref{keqk1}) is supplemented by the equation
for the retarded Green function. Following \cite{BM90} $A$ is the
algebraic function:
 \begin{eqnarray}\label{A}
 A (x,p) &=&\displaystyle \frac{\Gm (x,p)}{M^2 (x,p) + \Gm^2
(x,p) /4}+O(\partial_x^2),
\end{eqnarray}
 up to second order gradient terms.

As can be seen from Eqs. (\ref{F}), (\ref{G-def}) and
(\ref{Coll(kin)}), the gain rate $\Ldt$ differs from $F\Gamma/A$
only by corrections of the first order in the gradients
%
\begin{eqnarray}
\label{BM-subst} \Ldt = \Gm F/\A + C/\A = \Gm F/\A +O(\partial_x),
\end{eqnarray}
%
since $C \sim O(\partial_x)$.
This fact permits  to neglect the correction $O(\partial_x)$ to
 $\Ldt$ in the commutator term in the kinetic equation (\ref{keqk1}), as it leads to the
second-order in the gradient terms. Thus upon substitution
$\Ldt=\Gm F/\A$ in the commutator term \cite{BM90} one arrives at
the BM form of the kinetic equation,
%
\begin{eqnarray}
\label{keqk} \Do F (x,p) - \Pbr{\Gm\frac{F}{\A},\Re\Gr^R} &=& C
(x,p),
\end{eqnarray}
%
which is  equivalent to the KB form within the first-order
gradient approximation,  see \cite{IKV00} for details, (all terms
$\propto O(\partial_x^2)$ are now omitted).

In terms of the four-phase-space occupation functions $f(x,p)$ the
kinetic equation in the BM form can be rewritten as \cite{IKV00}
\begin{eqnarray}
\label{keqk2} \frac{A^2 \Gamma}{2}\left( \Do f(x,p) - \frac{M}{
\Gamma} \Pbr{\Gamma , f} \right) &=& C (x,p) .
\end{eqnarray}

Opening the Poisson brackets we arrive at equation
\begin{eqnarray}\label{BM}
\frac{A^2 \Gamma}{2}\left[ \left(v_{\mu}  - \frac{\partial
\Re\Sa^R}{\partial p^{\mu}}-\frac{M}{\Gamma} \frac{\partial
\Gamma}{\partial p^{\mu}} \right) \frac{\partial }{\partial
x_{\mu}}  +\left( \frac{\partial \Re\Sa^R}{\partial
x^{\mu}}+\frac{M}{\Gamma} \frac{\partial \Gamma}{\partial x^{\mu}}
\right) \frac{\partial }{\partial p_{\mu}}      \right]f =C.
\end{eqnarray}
With  the help of (\ref{F}), (\ref{G-def}) the collision term can
be expressed as
\begin{eqnarray}
\label{colGam}
 C=\Gamma_{\rm in}A -Af\Gamma.
\end{eqnarray}
In  the global thermal equilibrium
\begin{eqnarray}\label{Eq1}
f_{\rm eq.} =\frac{1}{\mbox{exp}[(p_0 -\mu )/T]\pm 1}.
\end{eqnarray}
This equation holds also in the local thermal equilibrium in
absence of the collective flow (at the velocity of the flow
$\vec{U}=0$),
\begin{eqnarray}\label{locEq10}
f_{\rm l.eq.}(\vec{U}=0) =\frac{1}{\mbox{exp}[(p_0 -\mu
(t,\vec{r}) )/T(t,\vec{r})]\pm 1},
\end{eqnarray}
provided  $T=T(t,\vec{r})$ and $\mu =\mu (t,\vec{r})$ are very
smooth functions of  $(t,\vec{r})$ (which vary on the space-time
scales much larger than the kinetic scale). With distribution
(\ref{locEq10}) the local part of the collision term vanishes;
$C^{\rm loc}=0$.

\subsection{Conservation  of charge and energy--momentum}
\label{Conservation-L}

Transport equation (\ref{keqk1}) weighted either with the charge
$e$, or with the 4-momentum $p^\nu $, integrated over momentum and
summed over internal degrees of freedom and particle species
($\Tr$) gives rise to the charge or energy--momentum conservation
laws with the density of the Noether 4-current and the Noether
energy--momentum tensor \cite{IKV00}
\begin{eqnarray}
\label{c-new-currentk}
j^{\mu} (x)
&=& e \mbox{Tr} \int \dpi{p}
\vu^{\mu}
\Fd (x,p), \\\label{E-M-new-tensork}
\Theta^{\mu\nu}(x)
&=&
\mbox{Tr} \int \dpi{p}
\vu^{\mu} p^{\nu} \Fd (x,p)
+ g^{\mu\nu}\left( {\cal E}^{\scr{int}}(x)-{\cal
E}^{\scr{pot}}(x)\right)\equiv \Theta^{\mu\nu}_{\rm
kin}+\Theta^{\mu\nu}_{\rm pot}.
\end{eqnarray}
 Here
%
\begin{eqnarray}
\label{eps-int} {\cal E}^{\scr{int}}(x)=\left<-\Lint(x)\right>
\end{eqnarray}
%
is the interaction energy density and
%
\begin{eqnarray}
\label{eps-potk} {\cal E}^{\scr{pot}}
=
\mbox{Tr} \int\dpi{p} \left[ \Re\Sa^R \Fd + \Re\Ga^R \Ldt
\right]= \mbox{Tr} \int\dpi{p} G_0^{-1}(p)Af
\end{eqnarray}
%
is the potential energy density. The first term in the squared
brackets in the first equality of (\ref{eps-potk}) complies with
quasiparticle expectations, namely mean potential times density,
the second term displays the role of fluctuations in the potential
energy density.

For specific interactions with the same number $\alpha$ of field
operators attached to any vertex of $\Lint$, one simply deduces
\cite{IKV99,IKV00,V08}
\begin{eqnarray}\label{int-spec}
{\cal E}^{\scr{int}}(x,p)=\frac{2}{\alpha}{\cal E}^{\scr{pot}}(x,p).
\end{eqnarray}

 For two-body non-relativistic interaction and for relativistic boson
 $\phi^4$
theory one gets $\alpha =4$. For a theory with two single-flavor
fermions interacting via one-flavor boson (with coupling
$\Psi_{\rm f}^{\dagger}\Psi_{\rm f} (\phi_{\rm b} +\phi_{\rm
b}^{\dagger})$) one obtains
 \begin{eqnarray}\label{twoFoneB}\epsilon_{\rm int}=\frac{2}{\alpha}(\epsilon_{\rm
pot}^{\rm f} +\epsilon_{\rm pot}^{\rm b} )=\frac{2}{\alpha_{\rm
f}} \epsilon_{\rm pot}^{\rm f} =\frac{2}{\alpha_{\rm b}}
\epsilon_{\rm pot}^{\rm b} , \quad \alpha =3, \,\, \alpha_{\rm
f}=2,\,\,\alpha_{\rm b}=1.
\end{eqnarray}
For a theory where two  fermions with different flavors interact
via one-flavor boson, one finds
 \begin{eqnarray}\label{twodFoneB}\epsilon_{\rm int}=2\epsilon_{\rm pot}^{{\rm f}_1}
 =2\epsilon_{\rm pot}^{{\rm f}_2}
=2\epsilon_{\rm pot}^{\rm b} .
\end{eqnarray}

For  relativistic particles  the energy-momentum tensor
(\ref{E-M-new-tensork})  is symmetric, i.e.
$\Theta^{\mu\nu}=\Theta^{\nu\mu}$. For non-relativistic particles
expression for the energy-momentum tensor is constructed from the
relativistic expression with the help of the expansion, where now
$p_0 \simeq m+p_0^{\rm n.-rel}$ and $|p_0^{\rm n.-rel}|\ll m$. As
follows from (\ref{E-M-new-tensork}), the non-relativistic value
$\Theta^{0i}$ only approximately coincides with $\Theta^{i0}$,
provided $v^\mu$ is given by Eq. (\ref{n-relvel}).

Ref. \cite{IKV00} has demonstrated that the conservation laws hold
in the form
\begin{eqnarray}\label{con-j}
\partial_{\mu}j^{\mu}(x)=0,\quad
\partial_{\mu}\Theta^{\mu\nu}(x)=0,
\end{eqnarray}
 provided all the self-energies are $\Phi$-derivable that we further
assume. The latter means that so called consistency conditions are
fulfilled:
%
\begin{equation}
\label{invarJk} \ii \mbox{Tr} \int \dpi{p} \left[ \Pbr{\Re\Sa^R,
\Fd} \Pbr{\Re\Ga^R,\Ldt } + C \right] =0,
\end{equation}
%

for the conserved current and
%
\begin{eqnarray}
\label{epsilon-invk}
\partial^{\nu}
\left( {\cal E}^{\scr{pot}} - {\cal E}^{\scr{int}} \right) =
\mbox{Tr} \int \frac{p^\nu \di^4 p}{(2\pi )^4} \left[
\Pbr{\Re\Sa^R, \Fd}
-
\Pbr{\Re\Ga^R,\Ldt } +C \right]
\end{eqnarray}
%
for the energy-momentum tensor.

All the properties of the KB-form of the kinetic  Eq.
(\ref{keqk1}) within a $\Phi$-derivable approximation also
transcribe to the BM-form, Eq. (\ref{keqk}), through the
substitution $\Ldt=\Gm F/\A$ in the consistency relations. However
in difference with the KB form of the kinetic equation
(\ref{keqk1}), where the Noether current (\ref{c-new-currentk})
and the Noether energy--momentum tensor (\ref{E-M-new-tensork})
are exactly conserved, with the BM form, Eq. (\ref{keqk}), the
conservation laws of the Noether current and the Noether
energy--momentum tensor are
 only approximately fulfilled (up to higher order gradients), see the proof in \cite{KIV01}.
 For systems in the thermal equilibrium, expressions for
thermodynamic quantities, being thermodynamically consistent with
expressions for the Noether current and energy-momentum, can be
found in \cite{V08}.

\section{Derivation of hydrodynamic equations for resonances}
\label{Hydro}

Below we follow the standard text-book procedure for the
derivation of the system of equations of the fluid dynamics
\cite{RR,LL79}, although with a specifics  that there is  no
strict dispersion relation between the energy and the momentum for
broad resonances.  To be specific and to easier clarify the
physical meaning of different terms entering the system of
equations of the fluid dynamics we further restrict ourselves by
consideration of non-relativistic fermions or bosons.

\subsection{Transformation between the laboratory and local rest frames and averaging procedure in 4-momentum
space}

Let in the laboratory frame (labeled by $'$) the given fluid
element moves with the  velocity $\vec U$. In the first
approximation we assume that, although  the system as a whole is
in non-equilibrium, it can be sub-divided on macroscopic but
physically small volumes, where the state can be considered as the
equilibrium one. Now let us consider $\vec{U}=\vec{U}(t,
\vec{r})$, as a smooth function of $(t, \vec{r})$, being
interpreted as the velocity of the center of inertia of a
physically small fluid volume, i.e. the local velocity of the
macroscopic motion of the fluid. Since $p^{\mu}x_{\mu}$ is
invariant, the energy and  momentum of the particle in the
laboratory frame and in the local rest frame are connected by
relations
\begin{eqnarray}\label{var}
p_{0}' \simeq p_0 +m\vec{v}\vec{U} +\frac{m \vec{U}^2}{2} ,\quad
\vec{p'}\simeq \vec{p}+m\vec{U}.
\end{eqnarray}
The spectral density in the local rest  frame is
\begin{eqnarray}\label{Spectrest}
A =A_{\rm l.eq.} (x, p_0 , \vec{p}^2),
\end{eqnarray}
and the local equilibrium distribution function $f_{\rm
l.eq.}(x,p_0)$ is given by Eq. (\ref{locEq10}), $p_0$ is
independent variable not related to the momentum.

 After the variable shift
(\ref{var}), the $\mbox{Tr}\int \dpi{p'} A'f' p_{i}'$ acquires the
form\footnote{We use Latin indices for the space-vector
components. }
\begin{eqnarray}\label{shift}
\mbox{Tr}\int \dpi{p'} A'f' p_{i}' = \rho U_i +\mbox{Tr} \int
\dpi{p} A f p_{i}=\rho U_i ,
\end{eqnarray}
where it is used that $Af$ may depend on $\vec{p}$ only in
$\vec{p}^2$ combination and thus
\begin{eqnarray}\label{def}
\mbox{Tr}\int \dpi{p} A f p_{i} =0,
\end{eqnarray}
demonstrating that the relative velocity of the chaotic thermal
motion in the local equilibrium in the rest frame is zero.

 The value
\begin{eqnarray}\label{rh}
\rho =m\mbox{Tr}\int \dpi{p}A f = m \overline{f}
\end{eqnarray}
 can be interpreted as
the mass density. Here we defined
\begin{eqnarray}\label{av}
\overline{\psi}=\mbox{Tr}\int \dpi{p} A\psi
\end{eqnarray}
for an   arbitrary function $\psi$.  Using  the sum-role
(\ref{sumA}) we can rewrite (\ref{av}) following  the standard
procedure of the averaging:
\begin{eqnarray}\label{av1}
\overline{\psi}= \mbox{Tr}\int \frac{\d^3 p}{(2\pi)^3}\left[\int
\frac{\d {p_0}}{2\pi} A\psi /\int \frac{\d {p_0}}{2\pi}A\right] .
\end{eqnarray}

Note that in the quasiparticle approximation,
\begin{eqnarray}\label{qpSum}
A_{\rm q.p.} (p_0 ,\vec{p}^2)
=
Z_0(2\pi)\delta \left(p_0 - \epsilon_p \right),\quad
Z_0=\left[v_0-\frac{\partial \Re\Sa^R}{\partial p_{0}}
\right]^{-1},
\end{eqnarray}
where $\epsilon_p$ is the root of the relation $M(\Gamma
\rightarrow 0)=0$,
 the sum-rule
(\ref{sumA}) is not satisfied, and (\ref{av}) and (\ref{av1}) are
not equivalent. With (\ref{qpSum}) Eq. (\ref{av}) yields
\begin{eqnarray}\label{av2}
\mbox{Tr}\int \dpi{p} A_{\rm q.p.}\psi=  \mbox{Tr}\int Z_0({p_0 =
\epsilon_p }) \psi \left[\epsilon (\vec{p}),
\vec{p}\right]\frac{\di^3 p}{(2\pi)^3}.
\end{eqnarray}
We also point out   that the quasiparticle effective 4-current and
the kinetic term in the energy-momentum tensor
\begin{eqnarray}
\label{c-new-currentkQP} j^{\mu}_{\rm q.p.}  &=& e \mbox{Tr} \int
\dpi{p} \left(v^{\mu}  - \frac{\partial \Re\Sa^R}{\partial
p_{\mu}}\right) A_{\rm q.p.} f , \\ \label{E-M-new-tensorkQP}
(\Theta^{\rm kin}_{\rm q.p.})^{\mu\nu} &=& \mbox{Tr} \int \dpi{p}
\left(v^{\mu}  - \frac{\partial \Re\Sa^R}{\partial p_{\mu}}\right)
p^{\nu} A_{\rm q.p.} f
\end{eqnarray}
 differ from the Noether ones by the
presence of the extra quasiparticle normalization factors
$v_{\mu}\rightarrow  \left(v_{\mu}  - \frac{\partial
\Re\Sa^R}{\partial p^{\mu}}\right)$ arising  from the interaction
terms. Using (\ref{qpSum}) we see that the normalization factor is
cancelled out in the definition of the density $\rho$ and its
presence results in the appearance of
 the effective mass instead of the bare mass in the expression for $\vec{j}$.
Further to describe resonances we use the definition (\ref{av}),
rather than (\ref{av1}), and the Noether quantities for the
current and the energy-momentum tensor.

\subsection{Continuity equation}

Presenting (\ref{c-new-currentk}) in the laboratory frame, then
doing the variable shift (\ref{var}) and using first Eq.
(\ref{con-j}) and Eq. (\ref{def})
we arrive at the continuity
equation
\begin{eqnarray}\label{contin}
\partial_t \rho +\mbox{div}(\rho\vec{U} )=0,
\end{eqnarray}
which has the standard form in these variables. The value
$\vec{j}=\rho \vec{U}$ is the density of the 3-mass-flow. In the
local rest frame, where $\vec{U}=0$, Eq. (\ref{contin}) is
rewritten as
\begin{eqnarray}\label{CONTINLOC}
\partial_t \rho +\rho\mbox{div}\vec{U} =0.
\end{eqnarray}
The continuity equation in the integral form is presented in
Appendix A (see Eq. (\ref{contin-int}) there).

\subsection{Momentum flow. Navier-Stokes equation}

Taking $\nu =i$ ($i=1,2$ or 3) component in Eq.
(\ref{E-M-new-tensork}) for the energy-momentum tensor, we present
the second Eq. (\ref{con-j}) as \footnote{Since further we deal
with non-relativistic dynamics, for spatial components we use
ordinary 3-dimensional notations, e.g., $x_k$ means
$(x^1,x^2,x^3)$.}
\begin{eqnarray}\label{mom}
\frac{\partial (\overline{mf' v_{i}' }) }{\partial t}+\frac{\partial
(\overline{m
    v_{i}' v_{k}' f'})}{\partial x_k}=(F_{i}^{\rm {int}})',
\end{eqnarray}
where $\vec{F}^{\rm int} =-\nabla ({\cal E}^{\scr{pot}}-{\cal
E}^{\scr{int}})$ can be interpreted as an internal force. After
the variable shift (\ref{var}), this equation is rewritten as
\begin{eqnarray}\label{mom1}
\frac{\partial (\rho U_i )}{\partial t} +\frac{\partial
(\overline{m f v_i v_k })}
{\partial
x_k}+\frac{\partial (\rho U_i U_k )}
{\partial
x_k}=F_i^{\rm int},
\end{eqnarray}
where we used equation
\begin{eqnarray}\label{avv}
\mbox{Tr}\int \dpi{p'} A'f' p_{i}' p_{k}' =\mbox{Tr}
\int \dpi{p} Af p_{i} p_{k} +
m^2 U_i U_k \mbox{Tr}\int \dpi{p} Af,
\end{eqnarray}
being obtained with the help of Eq. (\ref{def}).

Using (\ref{E-M-new-tensork}) we introduce the pressure according
to the local-equilibrium relation
\begin{eqnarray}\label{pres}
P =\frac{1}{3}(\Theta^{11}+\Theta^{22}+\Theta^{33})_{\rm l.eq.}
=P_{\rm kin} +({\cal E}^{\scr{pot} }-{\cal E}^{\scr{int}})_{\rm
l.eq.} ,
\end{eqnarray}
with the  kinetic contribution
\begin{eqnarray}\label{Pkin}
P_{\rm kin} =\frac{2}{3}\mbox{Tr}\int \dpi{p} \epsilon_p^0 A_{\rm
l.eq.} f_{\rm l.eq.} =\frac{2}{3}\overline{\epsilon_p^0 f_{\rm
l.eq.}}, \quad \epsilon_p^0 =\frac{\vec{p}^2}{2m}.
\end{eqnarray}
We also introduce a symmetric tensor $\Pi_{ik}$ as
\begin{eqnarray}\label{p-pi}
\Pi_{ik}(\vec{U})=P_{\rm kin} \delta_{ik} - m \overline{f v_{i}
v_{k} },
\end{eqnarray}
and the vector
\begin{eqnarray}\label{Ldef}
L_{k} (\nabla T)  \equiv \Theta^{k 0}=\overline{v_{k}p_0 f },
\end{eqnarray}
the physical meaning of the latter is clarified below. These are
straightforward generalizations of the standard Boltzmann
expressions, which will allow us to derive equations of the fluid
dynamics.

In the local equilibrium state in the local rest frame $f_{\rm
l.eq.}$ is the function of $p_{0}$, $T(t,\vec{r})$ and $\mu
(t,\vec{r})$, where $T$ is the temperature and  $\mu$ is the
chemical potential, see Eq. (\ref{locEq10}). Thereby
$m\overline{f_{\rm l.eq.} v_{i} v_{k} }=\frac{m}{3}
\overline{f_{\rm l.eq.}v^{ 2}}\delta_{ik} =P_{\rm kin}
\delta_{ik}$ resulting in $\Pi_{ik} =0$ and also $L_k=0$. Using
(\ref{p-pi}) and (\ref{pres}),
 and replacing ${\cal
E}^{\scr{pot} }-{\cal E}^{\scr{int}}$ to $({\cal E}^{\scr{pot}
}-{\cal E}^{\scr{int}})_{\rm l.eq.}+ \delta ({\cal E}^{\scr{pot}
}-{\cal E}^{\scr{int}})$ we rewrite Eq. (\ref{mom1}) in the
standard form
\begin{eqnarray}\label{mom2}
\frac{\partial (\rho U_i )}{\partial t} +\frac{\partial (\rho U_i
U_k )} {\partial x_k}\simeq -\frac{\partial P} {\partial
x_i}+\frac{\partial \Pi_{ik}} {\partial x_k} +\delta F_i .
\end{eqnarray}
 Eq. (\ref{mom2})
represents the second Newton law for unit fluid volume. Integral
form of this law is presented in Appendix A (see Eq.
(\ref{integ-mom})), $\delta F_i = -\frac{\partial\delta ({\cal
E}^{\scr{pot} }-{\cal E}^{\scr{int}})}{\partial x_i}$ can be
interpreted as a  force existing only in non-equilibrium ($\delta
F_i$ is zero in the local equilibrium state since $\delta({\cal
E}^{\scr{pot} }-{\cal E}^{\scr{int}})=0$). On the other hand the
term $\delta F_i$ can be presented as
\begin{eqnarray}\label{force}
\delta F_i =\frac{\partial\delta\Pi_{ik}}{\partial x_k},\quad
\delta\Pi_{ik}=-\delta_{ik}\delta({\cal E}^{\scr{pot} }-{\cal
E}^{\scr{int}}).
\end{eqnarray}
Note that the term (\ref{force}) associated with the interaction
in non-equilibrium state is usually ignored in practical
calculations based on the Boltzmann kinetics, as a sub-leading
term in the weak coupling limit and for dilute systems, see Eq.
(5.22) of \cite{Jeon} and Eq. (2.1) of \cite{SR09}. However there
are no arguments to omit this term   for a strong coupling and for
dense systems.

 In non-equilibrium states in the first approximation
$\Pi_{ik}$ should be proportional to the projections of the
gradient of the components of the velocity vector, since in order
a viscous friction of the near-by layers to appear, the velocities
of the layers should be different. Thus the trace-less
($\Pi_{ik}^{(1)}$) and diagonal ($\Pi_{ik}^{(2)}$) parts of
$\Pi_{ik}=\Pi_{ik}^{(1)}+\Pi_{ik}^{(2)}$, see Appendix A for more
detail, yield
\begin{eqnarray}\label{eta-zeta}
\Pi_{ik}^{(1)}=\eta W_{ik}\equiv \eta \left(U_{ik}-\frac{2}{3}
\frac{\partial U_l}{\partial x_l}\delta_{ik}\right),\quad
\Pi_{ik}^{(2)}=\zeta \frac{\partial U_l}{\partial
x_l}\delta_{ik},\quad \delta \Pi_{ik}=\delta\zeta \frac{\partial
U_l}{\partial x_l}\delta_{ik},
\end{eqnarray}
where $U_{ik}= \frac{\partial U_i}{\partial x_k}+ \frac{\partial
U_k}{\partial x_i},$ $\eta$ is the coefficient of the shear
(first) viscosity and $\zeta +\delta\zeta$, of the bulk (second)
viscosity. With these definitions Eq. (\ref{p-pi}) can be
rewritten as
\begin{eqnarray}\label{delkin}
\delta\Theta_{ik}^{\rm kin}\equiv \Theta_{ik}^{\rm kin}-P_{\rm
kin}\delta_{ik}=-\zeta \frac{\partial U_l}{\partial
x_l}\delta_{ik} -\eta W_{ik}=
\Theta_{ik}-P\delta_{ik}-\delta_{ik}\delta ({\cal E}^{\scr{pot}
}-{\cal E}^{\scr{int}}),
\end{eqnarray}
where we also used Eqs. (\ref{E-M-new-tensork}) and (\ref{pres}).

 For
specific interactions with the same number $\alpha$ of field
operators attached to any vertex,  expression (\ref{force}) is
well defined by Eqs. (\ref{eps-potk}), (\ref{int-spec}). E.g., for
a theory with two single flavor fermions interacting with one
flavor boson one has $\alpha_{\rm f}=2$ and $\alpha_{\rm b}=1$
that yields then no contribution to the bulk viscosity of the
fermion sub-system, $\delta\zeta_{\rm f}=0$, but produces a
contribution for the boson sub-system. On the other hand, one can
use $\alpha =3$ that allows to redistribute the
potential--interaction energy terms between both sub-systems. This
example teaches us that {\em{ the quantity $\delta\zeta$ is not
uniquely defined for the sub-system of a multi-component system
although it is uniquely defined for the system as a whole.}}
Possibility of a re-grouping of the interaction--potential energy
between sub-systems may allow one to easier calculate $\delta\zeta
=\sum_{a}\delta\zeta_{a}$.  In many practically interesting
situations a broad resonance appears, as a consequence of the
interaction between other particle species. Those (other) particle
species in many cases acquire much smaller widths than the given
broad resonance and thereby they can be treated within the
quasiparticle approximation. Thus it is convenient to relate the
interaction--potential energy term to the quasiparticle species
retaining broad resonances as quasi-free, see \cite{V08}, since
calculation of kinetic coefficients is easier done for
quasiparticles.

With the help of Eqs. (\ref{eta-zeta}), (\ref{delkin}) we rewrite
Eq. (\ref{mom2}) precisely in the Navier-Stokes form
\begin{eqnarray}\label{nav-st}
\frac{\partial (\rho U_i )}{\partial t} +\frac{\partial (\rho U_i
U_k )} {\partial x_k}+\frac{\partial P} {\partial x_i}-\eta
\frac{\partial^2  U_i }{\partial x_k^2}-\left(\zeta +\delta\zeta
+\frac{1}{3}\eta \right) \frac{\partial^2  U_k }{\partial
x_i\partial x_k} =0 .
\end{eqnarray}

Now we may use an identity for an arbitrary function $\psi$:
\begin{eqnarray}\label{aux}
\rho \left(\frac{\partial \psi }{\partial t}+\frac{\partial \psi
}{\partial x_k} \frac{\partial x_k }{\partial t}\right)=
\frac{\partial (\psi \rho )}{\partial t} +\frac{\partial (\psi
\rho U_k ) }{\partial x_k}.
\end{eqnarray}
To derive this identity we used the continuity Eq. (\ref{contin})
and that $\frac{\partial x_k }{\partial t}=U_k$. With the help of
this identity Eq. (\ref{mom2}) can be rewritten as
\begin{eqnarray}\label{mom3}
\rho \frac{\partial  U_i }{\partial t} +\rho U_k \frac{\partial
U_i } {\partial x_k}=-\frac{\partial P} {\partial
x_i}+\frac{\partial (\Pi_{ik}+\delta\Pi_{ik})} {\partial x_k} ,
\end{eqnarray}
and as
\begin{eqnarray}\label{mom30}
\rho \frac{\partial  U_i }{\partial t} =-\frac{\partial P}
{\partial x_i}+\frac{\partial (\Pi_{ik}+\delta\Pi_{ik})} {\partial
x_k}
\end{eqnarray}
in the local rest frame ($\vec{U}=0$). The value
\begin{eqnarray}
\delta P_{\rm
n.eq.}&=&\frac{1}{3}\left(\Theta^{11}+\Theta^{22}+\Theta^{33}\right)_{\rm
n.eq.}-
\frac{1}{3}\left(\Theta^{11}+\Theta^{22}+\Theta^{33}\right)_{\rm
l.eq.} \nonumber\\&=&-\frac{1}{3} (\Pi_{ii}^{(2)}+\delta
\Pi_{ii})=-(\zeta +\delta \zeta)\mbox{div}\vec{U}
\end{eqnarray}
has the meaning of the correction to the pressure at small
deviations from the local equilibrium (see Eq. (\ref{nu1-int}) in
Appendix A). Since $\zeta +\delta \zeta$ should be positive, the
sign of the correction to the pressure depends on the sign of
$\mbox{div}\vec{U}$. So, on the stage of expansion of the fireball
in the heavy ion collision the non-equilibrium pressure is in
reality  smaller than the equilibrium one being used in the ideal
hydrodynamics. This means that {\em{  realistic equilibrium
equation of state might be stiffer than that allows to fit
experimental data within ideal hydrodynamical simulations.}}

 If one sets $\Pi_{ik}+\delta\Pi_{ik}=0$, one arrives at  the Euler equation for the compressible ideal
 liquid
\begin{eqnarray}\label{mom2Loc}
\rho\frac{\partial  U_i }{\partial t} =-\frac{\partial P}
{\partial x_i}.
\end{eqnarray}
 In  thermal equilibrium  $\Pi_{ik}+\delta\Pi_{ik} =0$, and the fluid of resonances is not viscous one
 (in spite of the production and absorption of resonances are
included). For non-equilibrium systems the Euler equation may hold
only approximately (provided $\eta$ and $\zeta$ are very small).

\subsection{Energy flow}

Taking $\nu =0$ component of the energy-momentum tensor
(\ref{E-M-new-tensork}) in the second Eq. (\ref{con-j})  we obtain
\begin{eqnarray}\label{en-comp}
\frac{\partial (\overline{v_0 p_{0}' f' }) }{\partial
t}+\frac{\partial (\overline{v_{k}' p_{0}'f'})}{\partial x_k}
=\partial_t ({\cal E}^{\scr{pot} }-{\cal E}^{\scr{int}})' .
\end{eqnarray}
Using (\ref{var}) and (\ref{p-pi}) we rewrite Eq. (\ref{en-comp})
as
\begin{eqnarray}\label{nu}
&&\frac{\partial (\overline{v_0 p_{0} f }) }{\partial t}+
\frac{\partial (\rho \vec{U}^2/2 ) }{\partial t}+ \frac{\partial
(U_k \overline{p_{0} f }) }{\partial x_k}+ \frac{\partial (\rho
\vec{U}^2 U_k /2 ) }{\partial x_k}\nonumber \\ &&=-\frac{\partial
L_k }{\partial x_k}+
\partial_t ({\cal E}^{\scr{pot} }-{\cal E}^{\scr{int}})-
\frac{\partial (U_k P_{{\rm kin}} ) }{\partial x_k}+
\frac{\partial (U_i \Pi_{ik} ) }{\partial x_k}.
\end{eqnarray}

As follows from  Eq. (\ref{E-M-new-tensork}),  $\partial_t ({\cal
E}^{\scr{pot} }-{\cal E}^{\scr{int}})= \frac{\partial
(\overline{v_0 p_{0} f }) }{\partial t}-\partial_t \cal{E}$,
$\cal{E}\equiv \Theta^{00}$. Then  with the help of Eq.
(\ref{pres}) we rewrite Eq. (\ref{nu})  as
\begin{eqnarray}\label{nu1}
&&\frac{\partial \cal{E} }{\partial t}+ \frac{\partial (\rho
\vec{U}^2/2 ) }{\partial t}+ \frac{\partial (U_k\cal{E}) }
{\partial x_k}+
\frac{\partial (\rho \vec{U}^2 U_k /2 ) }{\partial x_k}\nonumber \\
&&=-\frac{\partial L_k }{\partial x_k}
-\frac{\partial (U_k P
) }{\partial x_k}+ \frac{\partial (U_i (\Pi_{ik} +\delta\Pi_{ik}))
}{\partial x_k}.
\end{eqnarray}
This equation describes  change of the energy with  passage of
time. Integral form of this equation is presented in Appendix A
(see Eq. (\ref{nu1-int})).

In order to do Eq. (\ref{nu}) self-closed we need an expression
for $\vec{L}$. Let us exploit  the fact that the heat conductivity
exists only in non-equilibrium states. Indeed, for the existence
of the heat flow one needs a temperature gradient. Then in the
first approximation
\begin{eqnarray}\label{L}
L_k =-\kappa \frac{\partial T}{\partial x_k},
\end{eqnarray}
where $\kappa$ is the coefficient of the heat conductivity
depending on the properties of the matter.

Multiplying Eq. (\ref{mom2}) by $U_i$ and using that
\begin{eqnarray}\label{aux1}
U_i \left(\frac{\partial (\rho U_i ) }{\partial t}+ \frac{\partial
(\rho U_i U_k ) }{\partial x_k} \right)=\frac{1}{2} \left(
\frac{\partial (\rho \vec{U}^2 ) }{\partial t}+ \frac{\partial
(\rho \vec{U}^2 U_k  ) }{\partial x_k} \right)
\end{eqnarray}
we obtain
\begin{eqnarray}\label{mom5}
\frac{\partial (\rho \vec{U}^2 /2) }{\partial t}+ \frac{\partial
(\rho \vec{U}^2 U_k /2 ) }{\partial x_k} =U_i \frac{\partial
(\Pi_{ik}+\delta\Pi_{ik}-P\delta_{ik}) }{\partial x_k}.
\end{eqnarray}
The l.h.s. is the  l.h.s. of the standard continuity equation, now for the
kinetic energy of the fluid.
In the r.h.s. of this equation we may recognize the work of the surface forces.

Using Eq. (\ref{mom5}) from (\ref{nu1}) we find
\begin{eqnarray}\label{en-dif}
\frac{\partial \cal{E}}{\partial t}+ \frac{\partial (U_k
\cal{E})}{\partial x_k}=-\frac{\partial L_k} {\partial
x_k}+\frac{\partial U_i}{\partial
x_k}\left(\Pi_{ik}+\delta\Pi_{ik}-P\delta_{ik}\right).
\end{eqnarray}
Thereby, we  recovered the standard form of the equation of the
fluid dynamics describing the energy transport. In the frame
$\vec{U}=0$ in the local equilibrium we get
\begin{eqnarray}\label{en-difLoc}
\frac{\partial \cal{E}}{\partial t}=-\left( \cal{E}+P\right)
\frac{\partial U_k}{\partial x_k}.
\end{eqnarray}
Integral form of Eq. (\ref{en-dif}) is presented in Appendix A
(see Eq. (\ref{mom5-int})).

\subsection{Equation for evolution of the entropy}

In the local equilibrium we may use the standard equilibrium expression
\begin{eqnarray}\label{ent-eq}
\di \widetilde{S} =\frac{1}{T}\di \widetilde{ {\cal{E}}}
+\frac{1}{T} P \di \widetilde{V} =\frac{1}{T}\di
\widetilde{\cal{E}} -\frac{P}{T}\frac{\di \rho}{\rho^2},
\end{eqnarray}
where tilde means that the quantity is related to the unit mass of
the liquid. The  sources of the entropy, which violate this
relation have smallness of the second space-time gradient order
\cite{IKV00}.  Thus, $d\widetilde{S}=0$ in the first space-time
gradient order and  thereby
\begin{eqnarray}
\frac{\partial \left(\cal{E}V\right)}{\partial
V}=-P+O(\partial_x^2).
\end{eqnarray}
Note that in Eqs. (\ref{mom3}), (\ref{en-dif}) we kept second
gradient order terms $\propto \left(\frac{\partial U_i}{\partial
x_k}\right)^2$ together with the first gradient order ones. Here
we drop  second gradient order terms compared to the zero order
ones.

From (\ref{ent-eq}) we get two equations
\begin{eqnarray}
\frac{\partial \widetilde{S}}{\partial t}=\frac{1}{T}
\frac{\partial \widetilde{\cal{E}}}{\partial t}-\frac{P}{T\rho^2}
\frac{\partial \rho}{\partial t},\quad U_k \frac{\partial
\widetilde{S}}{\partial x_k}=\frac{U_k}{T} \frac{\partial
\widetilde{\cal{E}}}{\partial x_k}-\frac{PU_k}{T\rho^2}
\frac{\partial \rho}{\partial x_k}.
\end{eqnarray}
Summing up these two relations and then multiplying the result by
$\rho$ and using Eq. (\ref{aux}) we find
\begin{eqnarray}
\frac{\partial (\rho\widetilde{S})}{\partial t}+ \frac{\partial
(\rho\widetilde{S}U_k )}{\partial x_k} =\frac{1}{T}\left[
\frac{\partial (\rho \widetilde{\cal{E}})}{\partial t}
+\frac{\partial (\rho \widetilde{\cal{E}}U_k)}{\partial
x_k}\right] -\frac{P}{T\rho} \left( \frac{\partial \rho}{\partial
t}+U_k \frac{\partial \rho}{\partial x_k} \right) .
\end{eqnarray}
Using (\ref{en-dif}) in the first term in the r.h.s. and
(\ref{contin}) in the second term we obtain
\begin{eqnarray}\label{en-fl}
\frac{\partial (\rho\widetilde{S})}{\partial t}+ \frac{\partial
(\rho\widetilde{S}U_k )}{\partial x_k}
=\frac{1}{T}\left((\Pi_{ik}+\delta\Pi_{ik})\frac{\partial
U_k}{\partial x_i}- \frac{\partial L_k}{\partial x_k}\right) .
\end{eqnarray}
We see that, if the r.h.s. of this equation were zero, we would
get the continuity equation for the entropy. Thus the r.h.s. of
Eq. (\ref{en-fl}) is  the density of the entropy sources. Its
integral form is presented in Appendix A (see Eq.
(\ref{en-fl-1})). In the local rest frame, $\vec{U}=0$, Eq.
(\ref{en-fl})  can be rewritten with the help of the continuity
Eq. (\ref{CONTINLOC}) as
\begin{eqnarray}\label{en-flLoc}
\rho\frac{\partial \widetilde{S}}{\partial t}= \frac{1}{T}\left[
\frac{\eta}{2} U_{ik}^2 +(\zeta +\delta\zeta -\frac{2}{3}\eta )
\left(\frac{\partial U_i}{\partial x_k} \right)^2
+\frac{\partial}{\partial x_k }\left(\kappa\frac{\partial
T}{\partial x_k}\right) \right]>0.
\end{eqnarray}
Only for $\eta, \zeta+\delta\zeta, \kappa \rightarrow 0$, the
entropy is conserved,
\begin{eqnarray}\label{en-flLoc1}
\frac{\partial \widetilde{S}}{\partial t}= 0.
\end{eqnarray}

Explicit expression for the Markovian entropy flow can be found in
\cite{IKV00}. Also,  on example of the first three diagrams of the
$\Phi$-functional for fermions interacting via a two-body
potential, there was found the memory correction to the entropy in
the local equilibrium.

\section{Approximate solution of the generalized kinetic equation}

Obtained above equations of  fluid dynamics of  resonance matter
enter three kinetic coefficients $\eta$, $\zeta$ and $\kappa$
which should be found from the solution of the Kadanoff-Baym
equations. Kinetic coefficients can be derived assuming that
deviations of the state of the moving fluid from the local
equilibrium  are small. To reproduce $\Pi_{ik}$ in accordance with
Eq.  (\ref{p-pi}) we need to find the solution $f$ of the
Kadanoff-Baym equations (\ref{keqk1}), or (\ref{keqk}), which are
equivalent  in the first gradient order.
 In
order to find slightly inhomogeneous solutions of these equations
with non-zero but small r.h.s. we  present
 \begin{eqnarray}\label{varf}
F = A_{\rm l.eq.}[f_{\rm l.eq.}](f_{\rm l.eq.}+\delta f)+\delta
A[\delta f] f_{\rm l.eq.}, \quad f= f_{\rm l.eq.}+\delta f,
\end{eqnarray}
where in the local equilibrium state in terms of the variables of
the laboratory frame $f_{\rm l.eq.}$ is given by
\begin{eqnarray}\label{locEq}
f_{\rm l.eq.} \simeq \frac{1}{\mbox{exp} [( p_0^{'} -\vec{p}'
\vec{U}-\mu (t,\vec{r}) )/T(t,\vec{r})]\pm 1},
\end{eqnarray}
and we suppose that $\vec{U} (t,\vec{r})$ is very small. By the
notation  $A_{\rm l.eq.}[f_{\rm l.eq.}]$ we stress that  $A_{\rm
l.eq.}$ depends functionally on $f_{\rm l.eq.}$.

As we have mentioned, the collision term \cite{IKV00} is
subdivided in two pieces, the local term $C^{\rm loc}$ and the
memory term $C^{\rm mem}$. Following (\ref{colGam}) one can
demonstrate that in the local equilibrium the local collision term
\begin{eqnarray}
C^{\rm loc}[f_{\rm l.eq.}]=A_{\rm l.eq.}[f_{\rm
l.eq.}]\left(\Gamma_{\rm in}[f_{\rm l.eq.}]-f_{\rm
l.eq.}\Gamma_{\rm l.eq.}[f_{\rm l.eq.}]\right)=0.
\end{eqnarray}
 To show this we  use that $C^{\rm
loc}[f_{\rm l.eq.}(p_0^{'}, \vec{p}^{'})]=C^{\rm loc}[f_{\rm
l.eq.}(p_0, \vec{p})]=0$ being obviously correct for $\vec{U},
\mu,T =const$. Since the local part of the collision term does not
depend on space-time gradients, the same is true for $\vec{U},
\mu,T $ being functions of $t,\vec{r}$.

Using (\ref{varf}) we find  variation of the collision integral
(\ref{Coll(kin)}):
\begin{eqnarray}\label{widG}\delta C^{\rm loc}=-A_{\rm l.eq.}\Gamma_{\rm l.eq.}\delta f
+A_{\rm l.eq.}\delta\Gamma^{\rm in}[f]  -A_{\rm l.eq.}f_{\rm
l.eq.}\delta\Gamma [ f]\equiv -A_{\rm
l.eq.}\widetilde{\Gamma}_{\rm l.eq.} \delta f.
\end{eqnarray}
Here there are terms containing $\delta \Gamma^{\rm in}[\delta f]$
and $\delta\Gamma [\delta f]$ which only implicitly depend on
$\delta f$. It is natural to expect that these dependencies  might
be weaker than that of the explicitly presented term. E.g. it is
so if one consideres the dynamics of the light particle admixture
in the medium of heavy particles. Then the light particle
self-energy can be considered as a function of the heavy particle
distributions, whereas a dependence on the distribution of the
light particle admixture can be neglected. Disregarding implicit
dependence of $\delta \Gamma^{\rm in}[\delta f]$ and $\delta\Gamma
[\delta f]$ on the phase-space integrals of $\delta f$ we get
\begin{eqnarray}\label{col-ans}
\delta C \simeq -A_{\rm l.eq.}\Gamma_{\rm l.eq.}\delta f.
\end{eqnarray}
Such an approximation, or better to say ansatz, is in spirit of
the known relaxation time approximation used in Boltzmann kinetics
to  describe near-equilibrium dynamics. Here all integrated
distributions are replaced by their local equilibrium values and
only not integrated one is allowed to have a non-equilibrium
variation. We will name this approximation, {\em{ local relaxation
time ansatz.}} In a particular case of the $\Phi$-functional
presented up to two vertices the proof of the validity of this
ansatz is presented in Appendix B. Also there we discuss a
difference of the local relaxation time ansatz and the global one.

Treating the kinetic equation in the BM form (\ref{BM}) in the
laboratory frame  perturbatively we express $\delta f$ through the
local equilibrium quantities
\begin{eqnarray}\label{BM1}
&&\left(\frac{A^2 \Gamma}{2}\left[ \left(v_{\mu}  - \frac{\partial
\Re\Sa^R}{\partial p^{\mu}}-\frac{M}{\Gamma} \frac{\partial
\Gamma}{\partial p^{\mu}} \right) \frac{\partial f}{\partial
x_{\mu}}  +\left( \frac{\partial \Re\Sa^R}{\partial
x^{\mu}}+\frac{M}{\Gamma} \frac{\partial \Gamma}{\partial x^{\mu}}
\right) \frac{\partial f}{\partial p_{\mu}} \right]\right)_{\rm
l.eq.}\nonumber\\ &&-C^{\rm mem}_{\rm l.eq.}=-A_{\rm
l.eq.}{\Gamma}_{\rm l.eq.} \delta f.
\end{eqnarray}
Here  index $"{\rm{l.eq.}"}$ indicates that all quantities are
taken at local equilibrium in  the laboratory frame, i.e.
expressed in $'$-variables. Also we may use that $\vec{U}$ is
small and put it zero after taking derivatives. Therefore one
should keep only time derivatives in the second term ($\propto
\frac{\partial f}{\partial p_{\mu}}$) in the l.h.s. of
(\ref{BM1}).

Since the collision term $C^{\rm mem}$ is of the first gradient
order, we may consider it as functional of $f_{\rm l.eq.}$. The
explicit form of the memory collision term depends on what
specific processes are studied. A specific example  is discussed
in Appendix C. The first non-vanishing diagram contributing to the
term $C^{\rm mem}$ has at least three vertices \cite{IKV00}. Thus
it contains an extra smallness at least for weak coupling.
 $\Gamma$ and
$\Re\Sa^R$ in the l.h.s. depend on $f_{\rm l.eq.}$ only
implicitly, i.e. through the phase-space integrals of $f_{\rm
l.eq.}$. This dependence is reflected in their dependence on
$\vec{U}(\vec{r},t)$, $\mu(\vec{r},t)$ and $T(\vec{r},t)$.
 Thus we arrive at equation for $\delta f$:
\begin{eqnarray}\label{BM2}
&&\left[\frac{A^2\Gamma }{2} \left(v_{\mu}  - \frac{\partial
\Re\Sa^R}{\partial p^{\mu}}-\frac{M}{\Gamma} \frac{\partial
\Gamma}{\partial p^{\mu}} \right)\right]_{\rm l.eq.}
\frac{\partial }{\partial x_{\mu}}  f_{\rm l.eq.}\nonumber\\
 &&+
\left[\frac{A^2\Gamma }{2}\left( \frac{\partial \Re\Sa^R
(p_0^{'}-\vec{p}^{'}\vec{U}, \vec{p}^{'}-m\vec{U},\mu,T)}{\partial
x^{0}}+\frac{M}{\Gamma} \frac{\partial
\Gamma(p_0^{'}-\vec{p}^{'}\vec{U},
\vec{p}^{'}-m\vec{U},\mu,T)}{\partial x^{0}} \right)\right]_{\rm
l.eq.} \nonumber\\ &&\times\frac{\partial f_{\rm l.eq.}}{\partial
p_{0}} -C^{\rm mem}_{\rm l.eq.}\simeq -A_{\rm l.eq.}{\Gamma}_{\rm
l.eq.}\delta f.
\end{eqnarray}

Below we use Eq. (\ref{BM2}) derived with the help of  the local
relaxation time ansatz (\ref{col-ans}). Using brief notations we
will suppress index $'$ which indicated the laboratory frame.
 Following (\ref{locEq}):
\begin{eqnarray}\label{fder}
\frac{\partial f_{\rm l.eq.}}{\partial t}&=&\frac{f_{\rm
l.eq.}(1\mp  f_{\rm l.eq.})}{T}\left[ \frac{(p_0
-\mu)}{T}\frac{\partial T}{\partial t}+ \vec{p}\frac{\partial
\vec{U} }{\partial t}+\frac{\partial \mu}{\partial
t}\right],\nonumber\\
 \nabla f_{\rm l.eq.}&=&\frac{f_{\rm
l.eq.}(1\mp  f_{\rm l.eq.})}{T}\left[ \frac{(p_0 -\mu)}{T}\nabla T
+ {p}_i \nabla {U}_i +\nabla \mu \right],
\end{eqnarray}
\begin{eqnarray}\label{fder1}
\frac{\partial f_{\rm l.eq.}}{\partial p_0 }=-\frac{f_{\rm
l.eq.}(1\mp  f_{\rm l.eq.})}{T},\quad \left(\frac{\partial f_{\rm
l.eq.}}{\partial \vec{p}}\right)_{\vec{U}=0}=0,
\end{eqnarray}
and  $\Sigma =\Sigma (p_0 -\vec{p}\vec{U}, \vec{p}-m\vec{U}, \mu
,T)$. After taking derivatives we everywhere  put $\vec{U}=0$.

Then we express $\delta f$ through the l.h.s. of Eq. (\ref{BM2})
as
\begin{eqnarray}\label{BMdeltaf}
\delta f &=&-\frac{A}{2}\frac{f(1\mp  f)}{T}\left[ \left(1  -
\frac{\partial \Re\Sa^R}{\partial p_{0}}-\frac{M}{\Gamma}
\frac{\partial \Gamma}{\partial p_{0}} \right)\frac{p_0 -\mu}{T}
-\frac{\partial \Re\Sa^R}{\partial T}-\frac{M}{\Gamma}
\frac{\partial \Gamma}{\partial T} \right]\frac{\partial
T}{\partial t}\nonumber \\ &-& \frac{A}{2}\frac{f(1\mp
f)}{T}\left[ 1  - \frac{\partial \Re\Sa^R}{\partial
p_{0}}-\frac{M}{\Gamma} \frac{\partial \Gamma}{\partial p_{0}}
-\frac{\partial \Re\Sa^R}{\partial \mu}-\frac{M}{\Gamma}
\frac{\partial \Gamma}{\partial \mu} \right]\frac{\partial
\mu}{\partial t}\nonumber \\
 &-&\frac{A}{2}\frac{f(1\mp
f)}{T} \vec{p}\frac{\partial \vec{U}}{\partial t} \nonumber \\
 &-&\frac{A}{2}\frac{f(1\mp
f)}{T}\left( \vec{v}+  \frac{\partial \Re\Sa^R}{\partial
\vec{p}}+\frac{M}{\Gamma} \frac{\partial \Gamma}{\partial \vec{p}}
 \right)\left(\frac{p_0 -\mu}{T}\nabla T +\nabla \mu +
 p_i \nabla U_i
 \right).
\end{eqnarray}
All quantities in the r.h.s.  are taken at the local equilibrium.
Not to complicate consideration we omitted contribution of $C^{\rm
mem}$. Demonstration how one can include contribution of  $C^{\rm
mem}$ is given in Appendix C.

\section{Kinetic coefficients}

To calculate shear viscosity we present \cite{LL79}
\begin{eqnarray}\label{deleta}
\delta f =T^{-1}f_{\rm l.eq.} g_{lh}U_{lh},
\end{eqnarray}
for $i\neq k$. From (\ref{p-pi}) one has
\begin{eqnarray}
\Pi_{ik}=-\frac{1}{T}\mbox{Tr}\int \frac{ d^4 p }{(2\pi)^4} v_{i}
p_{k} A_{\rm l.eq.} f_{\rm l.eq.} g_{lh} U_{lh} \equiv
\frac{1}{2}\eta_{iklh} U_{lh}.
\end{eqnarray}
The quantities  $\eta_{iklh}$ form  tensor of the rank 4,
symmetric in indices $ik$ and $lh$, being zero at the contraction
with respect to the pair $lh$. Since the fluid is isotropic, this
tensor is then expressed as:
$\eta_{iklh}=\eta[\delta_{il}\delta_{kh}+\delta_{ih}\delta_{kl}-\frac{2}{3}
\delta_{ik}\delta_{lh}]$. Thus we obtain
\begin{eqnarray}\label{et}
\eta =-\frac{1}{10T}\mbox{Tr}\int \frac{ d^4 p }{(2\pi)^4 } v_{l}
p_{h} g_{lh} f_{\rm l.eq.} A_{\rm l.eq.} .
\end{eqnarray}
In order to calculate the bulk viscosity we present
\begin{eqnarray}\label{delze1}
\delta f=T^{-1}f_{\rm l.eq.} g\, \mbox{div} \vec{U}.
\end{eqnarray}
Then we obtain
\begin{eqnarray}\label{ze}
\zeta =-\frac{1}{3T}\mbox{Tr}\int \frac{ d^4 p }{(2\pi)^4 }
p_{i}v_{i} g f_{\rm l.eq.} A_{\rm l.eq.} ,
\end{eqnarray}
\begin{eqnarray}\label{delze}
\delta \zeta =-\frac{(1-2/\alpha)}{T}\mbox{Tr}\int \frac{ d^4 p
}{(2\pi)^4 } (p_0 -m-\frac{p^2}{2m}) g f_{\rm l.eq.} A_{\rm l.eq.}
.
\end{eqnarray}
To derive (\ref{delze}) we used Eqs. (\ref{force}),
(\ref{eta-zeta}) and (\ref{eps-potk}), (\ref{int-spec}).
Performing variations following local relaxation time ansatz we
omitted variations $\delta A[\delta f]$ since $A$ depends on
$\delta f$ only implicitly. In the weak coupling limit and for the
low densities one has $\delta \zeta\ll \zeta$, cf. free resonance
case in \cite{V08}.

The energy flux is given as $L_i =\Theta^{i0} =\mbox{Tr}\int
\frac{ d^4 p}{(2\pi)^4}v_{i} p_0 A_{\rm l.eq.}f$, cf.
(\ref{Ldef}). We search
\begin{eqnarray}\label{delkap}
\delta f=g_i \frac{\partial T}{\partial x_i} f_{\rm l.eq.}.
\end{eqnarray}
Then using Eq. (\ref{L}) we find
\begin{eqnarray}\label{kap}
\kappa = -\frac{1}{3}\mbox{Tr}\int \frac{ d^4 p}{(2\pi)^4 } v_{i}
p_{0} g_i f_{\rm l.eq.} A_{\rm l.eq.} .
\end{eqnarray}

Substituting Eqs. (\ref{deleta}), (\ref{delze1}), (\ref{delkap})
into Eq. (\ref{BMdeltaf}) and using expressions (\ref{Ttmu}),
(\ref{PV}) derived in Appendix D we find\footnote{Further to
shorten notations we suppress index indicating local equilibrium.}
\begin{eqnarray}\label{gik}
g_{ik}= -\frac{A f(1\mp f)}{2}\frac{p_i
p_k}{m}\left[1+\frac{\partial \Re\Sa^R}{\partial
\epsilon_{p}^{0}}+\frac{M}{\Gamma} \frac{\partial \Gamma}{\partial
\epsilon_{p}^{0}}
 \right],
\end{eqnarray}
\begin{eqnarray}\label{gzeta}
g=-\frac{A (1\mp f)}{2}\widetilde{Z}_0^{-1} mI_{\zeta},
\end{eqnarray}
\begin{eqnarray}\label{gkap}
\vec{g}=-\vec{v}\frac{A (1\mp f)}{2T^2}\left(1+\frac{\partial
\Re\Sa^R}{\partial \epsilon_{p}^{0}}+\frac{M}{\Gamma}
\frac{\partial \Gamma}{\partial \epsilon_{p}^{0}}
 \right) \left( p_0 -\mu -T{S}\right),
\end{eqnarray}
where $S$ is the entropy per baryon,
\begin{eqnarray}\label{Ig}
 I_{\zeta} &=&
 \frac{\vec{p}^2}{3m m^*}-\left[ 1 -\widetilde{Z}_0 \left(
\frac{\partial \Re\Sa^R}{\partial \mu}+\frac{M}{\Gamma}
\frac{\partial \Gamma}{\partial \mu} \right)
\right]\left(\frac{\partial P}{\partial \rho}\right)_{\Theta_{00}}
\nonumber\\ &-& \left\{\frac{p_0}{m} -\widetilde{Z}_0 \left[
\frac{T}{m}\left(\frac{\partial \Re\Sa^R}{\partial
T}+\frac{M}{\Gamma} \frac{\partial \Gamma}{\partial
T}\right)+\frac{\mu}{m}\left(\frac{\partial \Re\Sa^R}{\partial
\mu}+\frac{M}{\Gamma} \frac{\partial \Gamma}{\partial
\mu}\right)\right]\right\}\left(\frac{\partial P}{\partial
\cal{E}}\right)_{n} .
\end{eqnarray}
Here we introduced the renormalization factor $\widetilde{Z}_0$
and  the ratio of the group velocity  to the phase  velocity:
\begin{eqnarray}\label{Z0}
\widetilde{Z}_0= \left(1 -\frac{\partial \Re\Sa^R}{\partial
p_0}-\frac{M}{\Gamma} \frac{\partial \Gamma}{\partial
p_0}\right)^{-1},\quad \frac{v_{\rm group}}{v_{\rm phase}}\equiv
\frac{m}{m^*}=\widetilde{Z}_0\left[1+\frac{\partial
\Re\Sa^R}{\partial \epsilon_{p}^{0}}+\frac{M}{\Gamma}
\frac{\partial \Gamma}{\partial \epsilon_{p}^{0}}
 \right],
\end{eqnarray}
$\vec{v}_{\rm phase}=\vec{p}/m$. The value $\vec{v}_{\rm
group}\rightarrow d \epsilon_{p}/d\vec{p}$ for $\Gamma \rightarrow
0$, as it follows from the dispersion relation (\ref{quas}). Then
the value $m^{*}$ for low momenta has the meaning of the Landau
(non-relativistic) effective mass ($v_{\rm group}\rightarrow
\vec{p}/m^{*}$). Deriving expression for the heat conductivity we
also used that following (\ref{PV}) the terms $\propto
\vec{v}\frac{\partial \vec{U}}{\partial t}$ appearing in
(\ref{BMdeltaf}) yield no contribution.

In order not to complicate consideration we omitted a contribution
of the memory collision term to the kinetic coefficients, which
form depends on the specific diagrams under consideration. Memory
term can be calculated following the line shown in Appendix C.

Finally we obtain the following expressions for the kinetic
coefficients:
\begin{eqnarray}\label{etafin}
\eta =\frac{1}{15}\mbox{Tr} \int \dpi{p} \frac{A^2
\Gamma}{2}\tau_{\rm rel}\frac{\vec{p}^4  f(1\mp f)}{ Tm m^{*}} ,
\end{eqnarray}
\begin{eqnarray}\label{zetafin}
\zeta =\frac{1}{3}\mbox{Tr} \int \dpi{p} \frac{A^2
\Gamma}{2}\tau_{\rm rel}\frac{\vec{p}^2  f(1\mp f)}{ T}I_{\zeta} ,
\end{eqnarray}
\begin{eqnarray}\label{zetafin}
\delta\zeta =(1-\frac{2}{\alpha})\mbox{Tr} \int \dpi{p} \frac{A^2
\Gamma}{2}\tau_{\rm rel}m(p_0 -m-\frac{\vec{p}^2}{2m})\frac{
f(1\mp f)}{ T}I_{\zeta} ,
\end{eqnarray}
\begin{eqnarray}\label{kappafin}
\kappa =\frac{1}{3}\mbox{Tr} \int \dpi{p} \frac{A^2
\Gamma}{2}\tau_{\rm rel}\frac{\vec{p}^2  f(1\mp f)}{ T}\frac{ p_0
(p_0 -\mu - T{S})}{m m^{*\,} T} ,
\end{eqnarray}
where
\begin{eqnarray}\label{taurel}
\tau_{\rm rel}=\widetilde{Z}_0^{-1} \Gamma^{-1}
\end{eqnarray}
has the meaning of a relaxation  time of the off-mass shell
particle. Often one determines heat conductivity through the
energy flux relative to the baryonic enthalpy, see \cite{Gavin}.
The shift to the Eckart frame results in quadratic expression
\begin{eqnarray}\label{kappafinE}
\kappa_{\rm E} =\frac{1}{3}\mbox{Tr} \int \dpi{p} \frac{A^2
\Gamma}{2}\tau_{\rm rel}\frac{\vec{p}^2  f(1\mp f)}{ T}\frac{ (p_0
-\mu - T{S})^2}{m m^* T} .
\end{eqnarray}

 Our expressions for the kinetic coefficients present
generalizations of expressions derived previously in different
limit cases. Ref. \cite{BS06} introduced expression for $\eta$ in
case of broad resonances  at assumption that $\Re\Sa^R$ and
$\Gamma$ are approximately constants and therefore their
derivatives are  zero. Thus our expression (\ref{etafin}) for
$\eta$ is the natural generalization of the result \cite{BS06}.
Ref. \cite{KTV09} conjectured expression for $\eta$ for resonances
in relativistic case  at the assumption that $\Sigma$ does not
depend on $p_0$ and $\vec{p}$ but may depend on $\mu$ and $T$.
Equation for $\eta$ used in \cite{KTV09} is the natural
generalization of that given in Ref. \cite{BS06}.

The quasiparticle limit is reproduced if one replaces $\Gamma
\rightarrow 0$ in the Green functions. For $\Gamma\rightarrow 0$
the value
\begin{eqnarray}\label{qpA}
\frac{A^2 \Gamma}{2}\rightarrow (2\pi)\delta (M).
\end{eqnarray}
 Note that doing this replacement one should treat
$\tau_{\rm rel}^{\rm q.p.}$ in (\ref{taurel}) as a finite value.
Here $\Gamma_{\rm q.p.}$ should be calculated with the help of the
quasiparticle Green functions.
 The quasiparticle  kinetic coefficients are obtained, if one  uses
 expression for the  energy-momentum tensor
 (\ref{E-M-new-tensorkQP}) instead of
(\ref{E-M-new-tensork}). Since $\eta$ and $\zeta$ are derived from
expression for the $(\Theta^{\rm kin}_{\rm q.p.})^{ik}$ there
appears extra factor $Z^{-1}_{\rm visc.} = 1+\frac{\partial
\mbox{Re}\Sigma^R}{\partial \epsilon_{p}^0}$. The quasiparticle
expression for $\kappa$ follows from the $(\Theta^{\rm kin}_{\rm
q.p.})^{0i}$. Therefore there arises extra factor $Z^{-1}_{\rm
heat}=Z_0^{-1}=1 - \frac{\partial \mbox{Re}\Sigma^R}{\partial
p_0}$.
 After  multiplication of expressions for the viscosities by $Z^{-1}_{\rm visc.}$ and the replacement
$Z^{-1}_{\rm visc.}{A^2 \Gamma}/{2}\rightarrow (m/m^{* })\delta
(p_0 -\epsilon_p )$
 and after   multiplication of expression for the heat conductivity
 by $Z^{-1}_{\rm heat}$ and subsequent replacement     $Z^{-1}_{\rm heat}{A^2 \Gamma}/{2}\rightarrow
\delta (p_0 -\epsilon_p )$, where $\epsilon_p$ is the solution of
the dispersion equation
\begin{eqnarray}\label{qpdisp}
p_0 -m-\epsilon_p^0 -\Re\Sa^R (p_0,\vec{p}) =0,
\end{eqnarray}
 cf. Appendix E,
we arrive at appropriate expressions for the  quasiparticle
kinetic coefficients:
\begin{eqnarray}\label{etafinQP}
\eta^{\rm q.p.} =\frac{1}{15}\mbox{Tr} \int \frac{\d^3
p}{(2\pi)^3} \tau^{\rm q.p.} _{\rm rel}\frac{\vec{p}^4  f(1\mp
f)}{ Tm^{*\,2}_{\rm q.p.}},
\end{eqnarray}

\begin{eqnarray}\label{zetafinQP}
\zeta^{\rm q.p.} =\frac{1}{3}\mbox{Tr}\int \frac{\d^3 p}{(2\pi)^3}
\tau_{\rm rel}^{\rm q.p.}\frac{\vec{p}^2  f(1\mp f)}{
T}I_{\zeta}^{\rm q.p.} ,
\end{eqnarray}

\begin{eqnarray}\label{kappafinQP}
\kappa^{\rm q.p.}_{\rm E} =\frac{1}{3}\mbox{Tr}\int \frac{\d^3
p}{(2\pi)^3} \tau^{\rm q.p.} _{\rm rel} \frac{\vec{p}^2  f(1\mp
f)}{ T}\frac{ (\epsilon_{p} -\mu - T{S})^2}{m m^{*}_{\rm q.p.}T} ,
\end{eqnarray}
where  $m^{*}=Z^{-1}_{0}Z_{\rm visc.}$, $\tau_{\rm rel}^{\rm
q.p.}={Z}_{0}^{-1} \Gamma^{-1}_{\rm q.p.}$ and
\begin{eqnarray}\label{IgQP}
 I_{\zeta}^{\rm q.p.} &=& \frac{\vec{p}^2}{3 m^{*\,2}_{\rm q.p.}}-\frac{m}{m^{*}_{\rm q.p.}}\left[ 1 -{Z}_{0}
\frac{\partial \Re\Sa^R}{\partial \mu} \right]\left(\frac{\partial
P}{\partial \rho}\right)_{\Theta_{00}} \nonumber\\ &-&
\left\{\frac{\epsilon_p}{m^{*}_{\rm q.p.}} -{Z}_{0} \left[
\frac{T}{m^{*}_{\rm q.p.}}\frac{\partial \Re\Sa^R}{\partial
T}+\frac{\mu}{m^{*}_{\rm q.p.}}\frac{\partial \Re\Sa^R}{\partial
\mu}\right]\right\}\left(\frac{\partial P}{\partial
\cal{E}}\right)_{n}.
\end{eqnarray}
Obviously expressions for $\delta\zeta^{\rm q.p.}$ and
$\kappa^{\rm q.p.}$ have similar forms. However one should bear in
mind differences between definitions of purely quasiparticle terms
 (\ref{c-new-currentkQP}), (\ref{E-M-new-tensorkQP}) and exact
expressions (\ref{c-new-currentk}), (\ref{E-M-new-tensork}). Here
in order to avoid double counting it is important to fix, whether
one deals with the truncated kinetic equation for quasiparticles
or with the full generalized kinetic equation consistent with the
conservation of the Noether current and energy-momentum tensors.

 At additional
assumption that $\Sigma$ does not depend on $p_0$ and $\vec{p}$
but may depend on $\mu$ and $T$ our expressions  for $\eta^{\rm
q.p.}$ and $\zeta^{\rm q.p.}$ coincide with those expressions
derived previously in \cite{SR08,KTV09} (after one re-writes the
latter expressions in the non-relativistic limit). Additionally
setting $\mbox{Re}\Sigma \rightarrow 0$ we reproduce the
perturbation theory results, see \cite{ID}.

Finally we should note that  in the standard quasiparticle  Fermi
liquid theory one usually uses slightly different procedure to
obtain transport coefficients, see \cite{BaymPethick}. Let us
formulate the corresponding generalization of this procedure to
the case of finite mass-widths. Let us use an ansatz that
\begin{eqnarray}\label{AKhAnsatz}
F=A_{\rm l.eq.}[f](f_{\rm l.eq.}+\delta f ),
\end{eqnarray}
where as above $\delta f =f-f_{\rm l.eq.}$ but with $A[f]$ being
functional of non-equilibrium $f$ rather than of $f_{\rm l.eq.}$.
One can prove, see Appendix F, that replacing $F=A_{\rm
l.eq.}[f]f_{\rm l.eq.}$ in the local collision term yields zero as
it was  for $F=A_{\rm l.eq.}[f_{\rm l.eq.}]f_{\rm l.eq.}$. Now
finding terms $\propto\delta f$  we should not vary $A$,
$\mbox{Re}\Sigma^R $ and $\Gamma $, since they depend on exact
$f$. Thus instead of (\ref{BM1}) we  get
\begin{eqnarray}\label{BM5}
&&\frac{A^2\Gamma }{2}\left[ \left(v_{\mu}  - \frac{\partial
\Re\Sa^R}{\partial p^{\mu}}-\frac{M}{\Gamma} \frac{\partial
\Gamma}{\partial p^{\mu}} \right) \frac{\partial f_{\rm
l.eq.}}{\partial x_{\mu}}  +\left( \frac{\partial
\Re\Sa^R}{\partial x^{\mu}}+\frac{M}{\Gamma} \frac{\partial
\Gamma}{\partial x^{\mu}} \right) \frac{\partial f_{\rm
l.eq.}}{\partial p_{\mu}} \right] \nonumber \\&&-C^{\rm
mem}[A,f_{\rm l.eq.}]=- A\Gamma\delta f.
\end{eqnarray}
Although the r.h.s. of this kinetic equation looks more simple
than that with the collision term (\ref{widG}), now all quantities
except $f_{\rm l.eq.}$ depend on unknown function $f$. In the
quasiparticle approximation this unknown dependence on
non-equilibrium distribution $f$ is hidden in the values of the
quasiparticle energies.
 However this is only an
apparent simplification since  the effective mass may depend on
$f$. Simplifying one often  ignores this dependence. Disregarding
implicit dependence of $m^{*}[f]$ is in the spirit of the local
relaxation time ansatz. Neglecting all implicit (functional)
dependencies on $f$ we actually do the same approximations as we
have done above within local relaxation time ansatz.

\section{Concluding remarks}

In conclusion, starting with expressions for the Noether current
and the energy-momentum tensor, as they follow from the gradient
expanded Kadanoff-Baym equations, we derived equations describing
the fluid dynamics of the non-relativistic system  of resonances
(particles with non-zero mass-widths). These equations, being
expressed in appropriate variables, have the same form as standard
equations of the fluid dynamics. The kinetic coefficients, the
shear and bulk viscosities and the heat conductivity, are
presented in terms of the self-energy functions and can be used
beyond the scope of the ordinary quasiparticle approximation. We
used  a local relaxation time ansatz to get explicit expressions
(in the Boltzmann kinetics a similar ansatz is called "the
relaxation time approximation"). We found a contribution of the
interaction--potential energy to the bulk viscosity and discussed
specifics of its interpretation for multi-component systems. We
also demonstrated how one can include memory contributions.
 Finally, we  discussed specifics of the quasiparticle
limit.

 {\bf
Acknowledgements} \vspace*{5mm}

I am   grateful to Y.B.~Ivanov, A.S. Khvorostukhin,
E.E.~Kolomeitsev, V.V. Skokov and V.D. Toneev for numerous
discussions. This work was supported in part by the DFG grant WA
431/8-1.

\vspace*{5mm}

{\bf{Appendix A. Equations of fluid dynamics in the integral
form}}

 Integrating the
continuity Eq. (\ref{contin}) over a fixed volume $V$ and using
the Gauss theorem one obtains
\begin{eqnarray}\label{contin-int}
\frac{\di \int \rho \di \vec{r}}{\di t}  =-m\int \di \sigma_k j_k
,
\end{eqnarray}
clearly showing that the change of the mass $\int \rho \di
\vec{r}$ in the given volume is determined by the mass-flow
through the surface. In relativistic problems particles and
anti-particles may annihilate in collisions producing  radiation
which should be taken into account. In non-relativistic systems
resonances can be absorbed and produced in particle collisions.
However their mass is assumed to be  redistributed in accordance
with the continuity equation, i.e. a small possible contribution
of the radiation is ignored.

  Integrating both parts of Eq. (\ref{mom2}) over a fixed
volume $V$ with the help of the Gauss theorem we obtain
\begin{eqnarray}\label{integ-mom}
\frac{\di (\int \rho U_i \di \vec{r})}{\di t}+\int \di \sigma_k
\rho U_i U_k
=
\int \di \sigma_k (\Pi_{ik}+\delta\Pi_{ik} -P \delta_{ik} ).
\end{eqnarray}
The l.h.s. is the change of the momentum in the volume $V$ due to
the change of the momentum at fixed point (first term) and due to
the fluid motion through the surface (second term). The r.h.s. is
the surface force $(\Pi_{ik}+\delta\Pi_{ik} -P \delta_{ik} )\di
\sigma_k$. In the local equilibrium $\Pi_{ik}+\delta\Pi_{ik}=0$
and the surface force is perpendicular to the element $\di
\sigma$, being equal to $-P\di \sigma$. In non-equilibrium states
there appears a force $\di \widetilde{F}_i
=(\Pi_{ik}+\delta\Pi_{ik}) \di \sigma_k$, acting on the element of
the surface square $\di \sigma_k$ and having tangential components
($i\neq k$). These components are associated with the viscous
friction. In order to explicitly separate them one presents
$\Pi_{ik}$ as the sum of two pieces,
$\Pi_{ik}=\Pi_{ik}^{(1)}+\Pi_{ik}^{(2)}$, where $\Pi_{ik}^{(1)}$
is the traceless tensor $\Pi_{ik}^{(1)}=K_{ik}-\frac{1}{3}
K\delta_{ik}$, $\mbox{Tr}\Pi_{ik}^{(1)}=0$, $K=\mbox{Tr}K_{ik}$,
and $\Pi_{ik}^{(2)}=N\delta_{ik}$, $\delta\Pi_{ik}=\delta N
\delta_{ik}$. The force due to the diagonal part, $\di
F^{(2)}=(\Pi_{ik}^{(2)}+\delta\Pi_{ik}) \di \sigma_k =(N+\delta N)
\di \sigma_i$, is orthogonal to the surface element $\di
\vec{\sigma}$ and has no tangential components. The traceless part
$\Pi_{ik}^{(1)}$ results in an additional contribution to the
non-equilibrium pressure $\di F^{(1)}_i n_i =\Pi_{ik}^{(1)} n_i
\di \sigma_k =(K_{ik}-\frac{1}{3}K\delta_{ik})n_i \di \sigma_k$
and also in tangential forces of viscous friction. The
$\Pi_{ik}+\delta\Pi_{ik}$ can be called the viscous stress tensor.
The term $\Pi_{ik}$ is reproduced from the Boltzmann kinetic
equation, whereas the term $\delta\Pi_{ik}$ does not arise there.

Integrating  Eq. (\ref{nu1}) over some fixed volume $V$ and using
the Gauss theorem we obtain
\begin{eqnarray}\label{nu1-int}
&&\frac{\di \left[ \int \di \vec{r} \left(\rho\vec{U} ^2/2
+\cal{E}\right)\right]}{\di t}+\int \di \sigma_k U_k\left(\rho
\vec{U}^2/2 + \cal{E} )\right)\nonumber \\ &&=\int
(\Pi_{ik}+\delta\Pi_{ik}- P \delta_{ik})U_i \di \sigma_k -\int L_k
\di \sigma_k .
\end{eqnarray}
In the l.h.s. one may recognize the change of the full energy in
the volume $V$ per unit time, as the consequence of the energy
change at fixed point and due to the particle motion through the
surface. This energy change consists of the kinetic energy and the
internal one. The first term in the r.h.s. is the work of the
surface forces, including the work of the pressure forces (the
equilibrium pressure, $P$, and the non-equilibrium one, $\Pi
=-\frac{1}{3} (\Pi_{ii}^{(2)}+\delta \Pi_{ii})$) and the work of
viscous friction forces. The last term in (\ref{nu1-int}) is the
flow of the vector $\vec{L}$ through the surface. It results in
the change of the energy in the volume $V$ even in absence of the
viscous friction. Thereby one can interpret this term as the heat
flow through the surface per unit time due to the heat
conductivity. Thus the vector $\vec{L}$ can be interpreted as the
vector of the heat flow. The value $\int \di \sigma_k
U_k(\frac{1}{2}\rho \vec{U}^2 + \cal{E} +P )$ has the meaning of
the heat content per unit volume.

After  integration of (\ref{en-dif}) over the volume $V$ we arrive
at
\begin{eqnarray}\label{mom5-int}
\frac{\di (\int \di \vec{r}\cal{E})}{\di t}+ \int \di \sigma_i U_i
\cal{E} =-\int L_i \di \sigma_i - \int P\frac{\partial U_i
}{\partial x_i}\di\vec{r} +\int \di
\vec{r}(\Pi_{ik}+\delta\Pi_{ik})\frac{\partial U_i }{\partial
x_k}.
\end{eqnarray}
This equation demonstrates that the reasons of the change of the
internal energy in the volume $V$ are the convection energy flow
through the surface (second term in the l.h.s.), the heat flow
(first term in the r.h.s.), the work of the pressure forces
(second term in the r.h.s.) and the work of the viscous forces
(third term in the r.h.s.). The viscosity always leads to a
decrease of the mechanical energy and to an increase of the
internal energy. Thereby the value
$(\Pi_{ik}+\delta\Pi_{ik})\frac{\partial U_i }{\partial x_k}$
should be positive in the non-equilibrium and zero in the
equilibrium. Using  Eqs. (\ref{eta-zeta}), we obtain
\begin{eqnarray}\label{et-zet}
(\Pi_{ik}+\delta\Pi_{ik})\frac{\partial U_i }{\partial
x_k}=\frac{1}{2}\eta U_{ik}^2 +\left( \zeta +\delta\zeta
-\frac{2}{3} \eta \right) \left( \frac{\partial U_k }{\partial
x_k}\right)^2 \geq 0.
\end{eqnarray}
Integrating both parts of Eq. (\ref{en-fl}) over the volume $V$
with the help of (\ref{et-zet}) we find
\begin{eqnarray}\label{en-fl-1}
\frac{\di (\int \rho\widetilde{S} \di \vec{r})}{\di t} &=&-\int
\rho\widetilde{S}U_k \di \sigma_k -\int \frac{L_k}{T}\di \sigma_k
\nonumber\\&+&\int\di \vec{r} \frac{1}{T}\left[ \frac{\eta}{2}
U_{ik}^2 +(\zeta +\delta\zeta -\frac{2}{3}\eta )
\left(\frac{\partial U_i}{\partial x_k} \right)^2\right] .
\end{eqnarray}
Thus, the change of the entropy consists of three parts. First
term in the r.h.s. is the convection entropy flow due to the
entropy transfer together with the fluid. Second term is the
consequence of the heat conductivity, being the heat flow, in
accordance with equation $\di \widetilde{S}=\di Q/T$. This flow
can be as positive as negative in dependence of the direction of
the vector $\vec{L}$ (of the temperature gradient). Third term
describes the appearance of the entropy due to the viscous
friction, being non-negative in agreement with the $H$ theorem.
This contribution is proportional to the  velocity derivatives
squared in agreement with the above assumption (\ref{ent-eq}) of
the quasi-equilibrium.

In the closed system there is no convection flow and the heat flow
(with heat isolated walls), and the entropy grows owing to the
viscous friction. If the system is in the local equilibrium, the
second and third terms in the r.h.s. of Eq. (\ref{en-fl-1}) being
zero and the entropy can only flow together with the fluid as the
whole. The entropy of the isolated system remains then  constant.

{\bf{Appendix B. Relaxation time ansatz.  Local relaxation time
ansatz and its check on a simple example. }}

{\bf{Generalized kinetic equation within relaxation time ansatz.}}
Let us first show how one may arrive at the generalized kinetic
equation within the relaxation time approximation, similar to what
is done in ordinary Boltzmann kinetics. Expanding all quantities
now {\em{ near the global equilibrium state}}, rather than near
the local equilibrium one, one obtains
\begin{eqnarray}\label{BM3}
&&\left[ \frac{A^2 \Gamma}{2} \left(v_{\mu}  - \frac{\partial
\Re\Sa^R}{\partial p^{\mu}}-\frac{M}{\Gamma} \frac{\partial
\Gamma}{\partial p^{\mu}} \right)\right]_{\rm eq.} \frac{\partial
\delta f }{\partial x_{\mu}}  \nonumber\\ &&+\left[\frac{A^2
\Gamma}{2}\right]_{\rm eq.}\left( \frac{\partial \delta \Re\Sa^R
[\delta f] }{\partial x^{0}}+\frac{M_{\rm eq.}}{\Gamma_{\rm eq.}}
\frac{\partial \delta \Gamma[\delta f]}{\partial x^{0}} \right)
\frac{\partial f_{\rm eq} }{\partial
 p_{0}} -\delta C^{\rm mem}=-A_{\rm eq}\widetilde{\Gamma}_{\rm eq} \delta
 f,
\end{eqnarray}
where now $\delta f =f-f_{\rm eq.}$. Here in the l.h.s. we
retained the terms which depend on $\delta f$ but vanish for the
equilibrium distribution. Then applying relaxation time ansatz,
i.e. dropping variations of all quantities, which depend on
$\delta f$ only implicitly,\footnote{Within the relaxation time
ansatz  one omits  $\delta C^{\rm mem}$. Note that in the local
equilibrium $C^{\rm mem}[f_{\rm l.eq.}]\neq 0$, see Eq.
(\ref{col-pi-delta}) in Appendix C, and one should keep this term
performing calculations within the local relaxation time ansatz.}
 we simplify Eq. (\ref{BM3}) as
\begin{eqnarray}\label{BM4}
\left[ \frac{A^2\Gamma }{2} \left(v_{\mu}  - \frac{\partial
\Re\Sa^R}{\partial p^{\mu}}-\frac{M}{\Gamma} \frac{\partial
\Gamma}{\partial p^{\mu}} \right)\right]_{\rm eq.} \frac{\partial
\delta f }{\partial x_{\mu}} \simeq -A_{\rm eq}\Gamma_{\rm
eq}\delta f,
\end{eqnarray}
 This equation can be treated
as {\em{ the generalized kinetic equation in, as usually called,
the relaxation time approximation.}}

{\bf{Local relaxation time ansatz.}} Now let us demonstrate the
validity of {\em{ the local relaxation time ansatz}}, when one
considers {\em{ small deviations from the local equilibrium
state.}} This is a weaker assumption than the relaxation time
ansatz just considered. Following the local relaxation time ansatz
one keeps the implicit dependencies of space-time derivatives of
the self-energies on $f_{\rm l.eq.}(x,p)$ but one drops their
implicit dependencies on $\delta f$.

To be specific consider a system of fermions interacting via a
two-body potential $V=V_0 \delta(x-y)$, and, for the sake of
simplicity, disregard its spin structure, by relating spin and
anti-symmetrization effects to a degeneracy factor $d$. For the
first two diagrams within the $\Phi$-derivable scheme (full Green
functions and free vertices), the
 self-energy becomes
\begin{eqnarray}\label{self2}
-\ii\left(\Se^{(1)} + \Se^{(2)}\right) = \pia \; + \;\;\pib \;
\end{eqnarray}
The self-energy here is presented up to two vertices. In this
approximation there are no memory effects and the collision term
$C^{\rm mem}=0$.

The collision term related to the second diagram (\ref{self2}) is
given by
\begin{eqnarray}
\label{C20} C^{(2)}  &=& d^2 \int \dpi{p_1} \dpi{p_2} \dpi{p_3}
\left|\;\; \wa{-} \;\;\right|^2 \nonumber\\[5mm] &\times &
\delta^4\left(p + p_1 - p_2 - p_3\right) \left( \F_2\F_3 \Ft\Ft_1
- \Ft_2\Ft_3 \F\F_1 \right).
\end{eqnarray}
In accordance with the local relaxation time ansatz, to find
$\delta C^{(2)}$ we vary only $F$ (and $\Ft$) and do not vary
$F_1$, $F_2$, $F_3$ since they are integrated. Also, varying $F$
we vary $f$ but do not vary $A$ since the latter quantity depends
on $f$ only through the integrals. Thus we find
\begin{eqnarray}
\label{deltaC20} \delta C^{(2)}  &=&- d^2  V_0^2 A(x,p) \delta
f(x,p) \int \dpi{p_1} \dpi{p_2} \dpi{p_3} \delta^4\left(p + p_1 -
p_2 - p_3\right)\nonumber\\
&\times & \left( \F_2\F_3 \Ft_1
 - \Ft_2\Ft_3 \F_1\right).
\end{eqnarray}
On the other hand, using  Eqs. (\ref{F}), (\ref{G-def}) and
opening the structure of the second diagram (\ref{self2})
contributing to the width we obtain
\begin{eqnarray}\label{Gam2}
\Gamma &=&\Gamma_{\rm out}-\Gamma_{\rm in}=\ii
\Sigma^{+-}-\ii\Sigma^{-+}\nonumber\\ &=&d^2 V_0^2\int \dpi{p_1}
\dpi{p_2} \dpi{p_3}  \delta^4\left(p + p_1 - p_2 - p_3\right)
\left( \F_2\F_3 \Ft_1
 - \Ft_2\Ft_3 \F_1\right).
\end{eqnarray}
Comparing (\ref{deltaC20}) and (\ref{widG}) with the help of Eq.
(\ref{Gam2})  we find that $\widetilde{\Gamma} =\Gamma$. Thus, on
this particular example we proved the validity of the local
relaxation time ansatz.

{\bf{Appendix C. Memory collision term with local equilibrium
distributions}}

Let us continue to study example of  a system of fermions
interacting via a two-body potential studied in Appendix B. Now
consider contribution of the collision term up to three vertices,
\cite{IKV00}.

For the first three diagrams within the $\Phi$-derivable scheme,
the  self-energy becomes
%
\begin{eqnarray}
\label{Pi-123} -\ii \Se &=& -\ii\left(\Se^{(1)} + \Se^{(2)} +
\Se^{(3)} \right) = \nonumber
\\
&=&  \pia \; + \;\;\pib \; + \;\;\pic
\end{eqnarray}
The local part of the collision term can  be presented in the form
%
\begin{eqnarray}
\label{C30} C^{(2)} + C_{\scr{loc}}^{(3)} &=& d^2 \int \dpi{p_1}
\dpi{p_2} \dpi{p_3} \left( \left|\;\; \wa{-} \;\;+
\!\!\!\!\!\wb{-} \;\;\right|^2 - \left|\!\!\!\!\!\wb-
\;\;\right|^2\right) \nonumber
\\[5mm]
&\times & \delta^4\left(p + p_1 - p_2 - p_3\right) \left( \F_2\F_3
\Ft\Ft_1 - \Ft_2\Ft_3 \F\F_1 \right),
\end{eqnarray}
%
where all the vertices in the off-shell scattering amplitudes are
of the same sign, say $"-"$ for definiteness.

Also the collision term contains a {\em non-local} (memory) part
due to the third diagram
%
\begin{eqnarray}
\label{col-pi-delta} &&C_{\scr{mem}}^{(3)}(x,p) = \left[
\left(\Se_{+-}^{(3)}\right)_{\scr{mem}}(x,p)\Gr^{-+}(x,p)
-
\Gr^{+-}(x,p) \left(\Se_{-+}^{(3)}\right)_{\scr{mem}}(x,p) \right]
\nonumber \\ &&= \frac{\ii}{2} \int \frac{\di^4 p'}{(2\pi)^4}
\frac{1}{d}\left[ \widetilde{\Loop}^{+-}(x;p'+p,p) -
\widetilde{\Loop}^{-+}(x;p'+p,p) \right] \Pbr{\Loop^{+-},
\Loop^{-+}}_{p' ,x}\, .
\end{eqnarray}
%
Here $\Loop^{jk}$ are the loops in  the Wigner representation,
%
\begin{eqnarray}\label{Loop-p}
\Loop^{jk}(x, p')= \int \frac{\di^4 p^{\prime\prime}}{(2\pi)^4}
\widetilde{\Loop}^{jk}(x;
p'+p^{\prime\prime},p^{\prime\prime})\equiv i V_0 L^{jk}_{\rm
B},\quad i,j=\{-,+\},
\end{eqnarray}
%
%
\begin{eqnarray}\label{Ltilde}
\widetilde{\Loop}^{jk}(x;p' +p^{\prime\prime}) = d \ii V_0
\ii\Gr^{jk}(x, p' +p^{\prime\prime})
\ii\Gr^{kj}(x,p^{\prime\prime}).
\end{eqnarray}
$L^{jk}_{\rm B}$ is the loop relativistic boson self-energy with
unit vertices. To calculate contribution of the memory term to the
kinetic coefficients we need $C^{\rm mem}[f_{\rm l.eq.}]$.

In the local equilibrium, as well as in the global equilibrium,
the Green functions and self-energies obey simple relations
\begin{eqnarray}
\label{Geq} \cal{G}^{ik} = \left(\begin{array}{ccc} \left[1\mp
f_{\rm l.eq.}\right]\cal{G}^R\pm f_{\rm l.eq.} \cal{G}^A && \pm
\ii f_{\rm l.eq.} \cal{A}\\[3mm] -\ii \left[1\mp f_{\rm
l.eq.}\right] \cal{A} && -\left[1\mp f_{\rm l.eq.}\right]
\cal{G}^A\mp f_{\rm l.eq.} \cal{G}^R
\end{array}\right),
\end{eqnarray}
where $\cal{G}^{ik}=G^{ik}_{\rm l.eq.}$ or $\Sigma^{ik}_{\rm
l.eq.}$ (in notation of \cite{IKV00}) and $\cal{A}=A$ or $\Gamma$,
respectively,  $\cal{G}^A =[\cal{G}^R]^{*}$.

Using these relations we express (\ref{col-pi-delta}) for the
local equilibrium distributions  as
\begin{eqnarray}
\label{col-pi-delta1} &&C_{\scr{mem, l.eq.}}^{(3)}(x,p) =-
\frac{A_{\rm l.eq.}^{\rm F}(x;p)}{2} \int \frac{\di^4
p'}{(2\pi)^4} V_0^{3} A_{\rm l.eq.}^{\rm F}(x;p'+p) \Gamma^{\rm
B}_{\rm l.eq.}(x,p') \nonumber\\ &&\times\left[ f^{\rm F}_{\rm
l.eq.}(x,p)-f^{\rm F}_{\rm l.eq.}(x,p'+p)\right]\Pbr{f^{\rm
B}_{\rm l.eq.}, \Gamma^{\rm B}_{\rm l.eq.}}_{p' ,x}\,\\
&&\equiv-{A_{\rm l.eq.}^{\rm F}(x;p)} \int \frac{\di^4
p'}{(2\pi)^4}K_{\rm l.eq.}(x;p' ,p) \Pbr{f^{\rm B}_{\rm l.eq.},
\Gamma^{\rm B}_{\rm l.eq.}}_{p' ,x}.\nonumber
\end{eqnarray}
Since in the local equilibrium state all quantities are assumed to
be known, further calculations of the memory contributions to the
kinetic coefficients are straightforward. What we are able to say
already without calculations is that in the weak coupling limit
$C_{\scr{mem, l.eq.}}^{(3)}$ is small ($\propto V_0^3$) and can be
neglected, thereby.

To further proceed we need to calculate $\Pbr{f^{\rm B}_{\rm
l.eq.}, \Gamma^{\rm B}_{\rm l.eq.}}_{p' ,x}$. Simplifying
notations we will suppress index "${\rm l.eq.}$". Using
(\ref{fder}), (\ref{fder1}) and (\ref{Ttmu}), (\ref{PV}) and that
$\Gamma^{\rm B}=\Gamma^{\rm B}[T(x,t)]$ and $\mu^{\rm B}=0$ we
find
\begin{eqnarray}\label{PBcalc}
\Pbr{f^{\rm B}, \Gamma^{\rm B}}_{p' ,x}=\frac{f^{\rm B}(1+f^{\rm
B})}{T}\left(T\frac{\partial\Gamma^{\rm B}}{\partial
T}+p_0^{'}\frac{\partial\Gamma^{\rm B}}{\partial
p_0^{'}}\right)\left(\frac{\partial P}{\partial
\cal{E}}\right)_{n}\mbox{div}\vec{U}.
\end{eqnarray}
We omitted terms $\propto \nabla (\vec{p}\vec{U})$ and $\propto
\vec{p}\nabla T$ since they do not contribute to
$C_{\scr{mem}}^{(3)}$ disappearing after angular integrations.
Thus the memory term contributes only to the bulk viscosity.

Following  (\ref{BM2}) one has $\delta f_{\rm mem}^{\rm
F}=C_{\scr{mem}}^{(3)}/(A^{\rm F}\Gamma^{\rm F})$ and from
(\ref{ze}) we obtain
\begin{eqnarray}\label{zemem}
&&\delta\zeta^{(3)}_{\rm mem}=\frac{1}{3\Gamma^{\rm
F}(x;p)}\int\frac{d\,\di^4 p}{(2\pi)^4} \frac{\di^4
p'}{(2\pi)^4}\frac{\vec{p}^2}{m} K_{\rm l.eq.}(x;p' ,p)
\frac{f^{\rm B}(x;p')[1+f^{\rm B}(x;p')]}{T}\nonumber\\
&&\times\left(T\frac{\partial\Gamma^{\rm B}(x;p')}{\partial
T}+p_0^{'}\frac{\partial\Gamma^{\rm B}(x;p')}{\partial
p_0^{'}}\right)\left(\frac{\partial P}{\partial
\cal{E}}\right)_{n}.
\end{eqnarray}

{\bf{Appendix D. Space-time dependence  of thermodynamical
quantities in the local equilibrium system}}

To find kinetic coefficients we use perturbative analysis
considering their contributions as small. Therefore we  exploit
equations of motion (\ref{CONTINLOC}), (\ref{mom2Loc}),
(\ref{en-difLoc}) and (\ref{en-flLoc1}). First we express pressure
in different variables
\begin{eqnarray}\label{pressureVar}
P=P(n,\cal{E}),\quad P=P(\mu , T),\quad P=P(n,T).
\end{eqnarray}
Then using first of these equations and also Eqs.
(\ref{CONTINLOC}) and  (\ref{en-difLoc}) we obtain
\begin{eqnarray}\frac{\partial P}{\partial t}=-\left\{\left(\frac{\partial P}{\partial
\cal{E}}\right)_{n}\left[ T\left(\frac{\partial P}{\partial
T}\right)_{\mu}+\mu\left(\frac{\partial P}{\partial
\mu}\right)_{T}\right]- \left(\frac{\partial P}{\partial
n}\right)_{\cal{E}}\left(\frac{\partial P}{\partial
\mu}\right)_{T}\right\}\mbox{div}{\vec{U}}.
\end{eqnarray}
On the other hand from the second Eq. (\ref{pressureVar}) we find
that
\begin{eqnarray}\label{Pt}
\frac{\partial P}{\partial t}=\left(\frac{\partial P}{\partial
T}\right)_{\mu}\frac{\partial T}{\partial t}+\left(\frac{\partial
P}{\partial \mu}\right)_{T}\frac{\partial \mu}{\partial t}.
\end{eqnarray}
With the help of the latter two expressions we obtain
\begin{eqnarray}\label{Ttmu}
\frac{\partial T}{\partial t}=-\left(\frac{\partial P}{\partial
\cal{E}}\right)_{n} T\mbox{div}{\vec{U}},\quad \frac{\partial
\mu}{\partial t}=-\left[ \mu\left(\frac{\partial P}{\partial
\cal{E}}\right)_{n} +\left(\frac{\partial P}{\partial
n}\right)_{\cal{E}}\right]\mbox{div}{\vec{U}}.
\end{eqnarray}
 With the help of the standard thermodynamic
relation,  Eq. (\ref{mom2Loc}) is rewritten as
\begin{eqnarray}\label{PV}
\frac{\partial \vec{U}}{\partial t} =-\frac{\nabla P }{\rho}
=-\frac{\nabla \mu }{m} -\frac{{S} \nabla T}{m} =0,
\end{eqnarray}
since in the local equilibrium $\nabla P=0$.  Here $S$ is the
entropy per baryon.

Expressing $\widetilde{S}=\widetilde{S}(T,P)$ from
(\ref{en-flLoc}) we find
\begin{eqnarray}\label{Tt}
\frac{\partial T}{\partial t}=-T c_P^{-1}\left(\frac{\partial
\widetilde{S}}{\partial P}\right)_{T}\frac{\partial P}{\partial
t}, \quad c_P =T\left(\frac{\partial \widetilde{S}}{\partial
T}\right)_{P}.
\end{eqnarray}
From the third Eq. (\ref{pressureVar}) and Eq. (\ref{CONTINLOC}):
\begin{eqnarray}\label{Pt1}
\frac{\partial P}{\partial t}=\left(\frac{\partial P}{\partial
n}\right)_{T}n\mbox{div}\vec{U}+\left(\frac{\partial P}{\partial
T}\right)_{n}\frac{\partial T}{\partial t}.
\end{eqnarray}
Using it and also  (\ref{Pt}) and  (\ref{Tt}) we obtain
\begin{eqnarray}\label{Pt-Tt}
\frac{\partial P}{\partial t}=-n\left(\frac{\partial P}{\partial
n}\right)_{T}\left[1+\frac{T}{c_p}\left(\frac{\partial
\widetilde{S}}{\partial P}\right)_{T}\left(\frac{\partial
P}{\partial T}\right)_{n}\right]^{-1}\mbox{div}{\vec{U}} ,
\end{eqnarray}
and thus
\begin{eqnarray}
\left(\frac{\partial P}{\partial \cal{E}}\right)_{n}=-\frac{
n}{c_P}\left(\frac{\partial \widetilde{S}}{\partial
P}\right)_{T}\left(\frac{\partial P}{\partial
n}\right)_{T}\left[1+\frac{T}{c_p}\left(\frac{\partial
\widetilde{S}}{\partial P}\right)_{T}\left(\frac{\partial
P}{\partial T}\right)_{n}\right]^{-1}.
\end{eqnarray}

{\bf{Appendix E. Dependence of the spectral function on its
arguments in the laboratory frame }}

 The spectral function (\ref{A}) depends on $p_0^{'}$ through
specific combinations:
\begin{eqnarray}
&&p_0^{'} -m-\epsilon_{p^{'}}^{ 0} -\mbox{Re}\Sigma^R_{\rm l.eq.}
(p_0^{'},\vec{p}^{'\,2};\mu,T)
 + \widetilde{\alpha} \vec{p}' \vec{U}+O(U^2), \quad \mbox{and}\\
&&\Gamma_{\rm l.eq.}^{'}=\Gamma_{\rm l.eq.}
(p_0^{'},\vec{p}^{'\,2};\mu,T)-\widetilde{\beta} \vec{p}'
\vec{U}+O(U^2),
\end{eqnarray}
where
\begin{eqnarray}\widetilde{\alpha} =\frac{\partial
\mbox{Re}\Sigma^R_{\rm l.eq.}}{\partial p_0^{'}}+\frac{\partial
\mbox{Re}\Sigma^R_{\rm l.eq.}}{\partial \epsilon_{p^{'}}^0},\quad
\widetilde{\beta}=\left(\frac{\partial \Gamma_{\rm
l.eq.}}{\partial p_0^{'}}+\frac{\partial \Gamma_{\rm
l.eq.}}{\partial \epsilon_{p^{'}}^0}\right).
\end{eqnarray}
We used that $p_0 -m-\epsilon_p^0 = p_0^{'} -m-\epsilon_{p^{'}}^{
0}+O(U^2)$ since $p_0\simeq p_0^{'}-\vec{p}^{'} \vec{U}$,
$\vec{p}\simeq \vec{p}^{'} -m\vec{U}$ and therefore $\epsilon_p^0
\simeq \epsilon_{p{'}}^0 -\vec{p}^{'}\vec{U}$. Thus
\begin{eqnarray}\label{Ashift}
A_{\rm l.eq.}^{'}\simeq A\left[ p_0^{'} -m-\epsilon_{p^{'}}^{ 0}
-\mbox{Re}\Sigma^R_{\rm l.eq.} (p_0^{'},\vec{p}^{'\,2};\mu,T)
 + \widetilde{\alpha} \vec{p}' \vec{U};\Gamma^{'}_{\rm l.eq.}
 (p_0^{'},\vec{p}^{'\,2};\mu,T)\right].
\end{eqnarray}
 The argument $p_0^{'} -m-\epsilon_{p^{'}}^{ 0}
-\mbox{Re}\Sigma^R_{\rm l.eq.} (p_0^{'},\vec{p}^{'\,2};\mu,T)
 + \widetilde{\alpha} \vec{p}' \vec{U}$ can be further expanded in $p_0^{'} -\epsilon_p^{'
 }$, where $\epsilon_p^{'}$ is the root of the
 dispersion relation
\begin{eqnarray}
\label{quas} p_0^{'} -m-\epsilon_{p^{'}}^{ 0}
-\mbox{Re}\Sigma^R_{\rm l.eq.} (p_0^{'},\vec{p}^{'\,2};\mu,T)=0,
\end{eqnarray}
which appears in the quasiparticle approximation. Note that in
difference with the quasiparticle dispersion relation, here we do
not assume that $\Gamma\rightarrow 0$.
 Then we obtain
\begin{eqnarray}
p_0^{'} -m-\epsilon_{p^{'}}^{ 0} -\mbox{Re}\Sigma^R_{\rm l.eq.}
(p_0^{'},\vec{p}^{'\,2};\mu,T) \simeq \left(1 - \frac{\partial
\mbox{Re}\Sigma^R_{\rm l.eq.}}{\partial
p_0^{'}}\right)_{p_0={\epsilon_p^{' }}}(
 p_0^{'} -\epsilon_p^{' }).
\end{eqnarray}
Thus
\begin{eqnarray}\label{vartil}
&&p_0^{'} -m-\epsilon_{p^{'}}^{ 0} -\mbox{Re}\Sigma^R_{\rm l.eq.}
(p_0^{'},\vec{p}^{'\,2};\mu,T)
 + \widetilde{\alpha} \vec{p}' \vec{U}\nonumber\\
 &&\simeq \left(1 - \frac{\partial \mbox{Re}\Sigma^R_{\rm
l.eq.}}{\partial p_0^{'}}\right)_{p_0^{'}={\epsilon_p^{' }}}\left(
 p_0^{'} -\epsilon_p^{' } +\bar{\alpha} \vec{p}' \vec{U}\right),
\end{eqnarray}
where
\begin{eqnarray}\label{alpha}\bar{\alpha} =\left[ 1 -
\frac{\partial \mbox{Re}\Sigma^R_{\rm l.eq.}}{\partial
p_0^{'}}\right]^{-1}_{p_0^{'}={\epsilon_p^{'
}}}\left[\frac{\partial \mbox{Re}\Sigma^R_{\rm l.eq.}}{\partial
p_0^{'}}+\frac{\partial \mbox{Re}\Sigma^R_{\rm l.eq.}}{\partial
\epsilon_{p^{'}}^0}\right].
\end{eqnarray}
and in all quantities  $p_0^{'}=\epsilon_p^{' }$. In case when the
dispersion equation has many roots ($\epsilon_{p, a}^{' }$), the
spectral function can be approximated by  the sum of the
corresponding terms.
 Note that the renormalization
coefficients $\frac{\partial \mbox{Re}\Sigma^R_{\rm
l.eq.}}{\partial p_0}$ and $ \frac{\partial \mbox{Re}\Sigma^R_{\rm
l.eq.}}{\partial \epsilon_{p}^0}$ are absent only in the mean
field approximation, when only tadpole diagrams are included.
Finally
\begin{eqnarray}\label{Ashift1}
A_{\rm l.eq.}^{'}\simeq A\left[ \left(1 - \frac{\partial
\mbox{Re}\Sigma^R_{\rm l.eq.}}{\partial
p_0^{'}}\right)_{p_0^{'}={\epsilon_p^{' }}}\left(
 p_0^{'} -\epsilon_p^{' } +\bar{\alpha} \vec{p}'
 \vec{U}\right);\Gamma_{\rm l.eq.}-\widetilde{\beta} \vec{p}' \vec{U}
 \right].
\end{eqnarray}
One can neglect the term $\propto \widetilde{\beta}$  since it is
$O(U\Gamma)$ provided both $U$ and $\Gamma$ are small. The
quasiparticle limit expression (for $\Gamma\rightarrow 0$) becomes
\begin{eqnarray}\label{Ashift1}
A_{\rm l.eq.}^{'}\simeq \left(1 - \frac{\partial
\mbox{Re}\Sigma^R_{\rm l.eq.}}{\partial
p_0^{'}}\right)^{-1}_{p_0^{'}={\epsilon_p^{' }}} (2\pi)\delta
\left(
 p_0^{'} -\epsilon_p^{' } +\bar{\alpha} \vec{p}'
 \vec{U}\right).
\end{eqnarray}
\\

{\bf{Appendix F. Local collision term for two presentations of
$F$}}

Here we will show that the collision term $C^{\rm loc}=0$ not only
for $F_{\rm l.eq.}$ but also for $F$ introduced by Eq.
(\ref{AKhAnsatz}). To do this consider  example of two-fermion
interaction via two-body potential given in Appendix B. Consider
three first diagrams (\ref{Pi-123}) with full Green functions and
free vertices. With (\ref{AKhAnsatz}) the local collision term
(\ref{Coll(kin)}) can be presented as
\begin{eqnarray}\label{colFermi}
 &&C^{(2)} + C_{\scr{loc}}^{(3)} = d^2 \int \dpi{p_1} \dpi{p_2}
\dpi{p_3} \left( \left|\;\; \wa{-} \;\;+ \!\!\!\!\!\wb{-}
\;\;\right|^2 - \left|\!\!\!\!\!\wb- \;\;\right|^2\right)
\nonumber
\\[5mm]
&&\times  \delta^4\left(p + p_1 - p_2 - p_3\right) A(x,p)A(x,p_1
)A(x,p_2 )A(x,p_3 )\\ &&\times\left[ f_{\rm l.eq.} (x,p_{2})f_{\rm
l.eq.}(x,p_{3})(1- f_{\rm l.eq.}(x,p_{1}))(1- f_{\rm
l.eq.}(x,p))\right.\nonumber\\&&\left. -(1- f_{\rm
l.eq.}(x,p_{2}))(1- f_{\rm l.eq.}(x,p_{3}))f_{\rm
l.eq.}(x,p_{1})f_{\rm l.eq.}(x,p)\right] .\nonumber
\end{eqnarray}
Local equilibrium distributions (\ref{locEq}) fulfill  relation
\begin{eqnarray}
\pm f_{\rm F/B}^{\rm l.eq.}(x,p+q)[1\mp f_{\rm F/B}^{\rm
l.eq.}(x,p)]=[f_{\rm F/B}^{\rm l.eq.}(x,p)\mp f_{\rm F/B}^{\rm
l.eq.}(x,p+q)]f_{\rm B/F}^{\rm l.eq.}(x,q)
\end{eqnarray}
for fermions and bosons (F/B). With the help of this relation we
may see that the term in the squared bracket in (\ref{colFermi})
is zero independent on the values of $A$. Thus $C^{\rm loc}=0$ for
both distributions $F=A[f_{\rm l.eq.}]f_{\rm l.eq.}$ and
$F=A[f]f_{\rm l.eq.}$.

\end{fmffile}

\end{document}

%% file: entr-mac.tex
\let\sectiontmp\section\let\subsectiontmp\subsection \let\appendixtmp\appendix
\def\section{\setcounter{equation}{0}\sectiontmp}
\def\subsection{\subsectiontmp}
\def\theequation{\arabic{section}.\arabic{equation}}
\def\appendix{\def\theequation{\Alph{section}.\arabic{equation}}\appendixtmp}
\input{graphs}

\newcommand{\di}{{\mathrm d}}
\newcommand{\Tr}{{\mathrm{Tr}}}
\newcommand{\ii}{{\mathrm i}}

\renewcommand{\and}{\quad{\mathrm{and}}\quad}

\renewcommand{\Re}{{\mathrm{Re}}}

\renewcommand{\Im}{{\mathrm{Im}}}

\def\scr#1{\mbox{\scriptsize #1}}

\def\vec#1{\mbox{\boldmath $#1$}}
\newcommand{\dpi}[1]{\frac{\di^4 #1}{(2\pi)^4}}                
\newcommand{\Pbr}[1]{\left\{#1\right\}}                    
\newlength{\charwidth}

\newcommand{\lap}%
{\raisebox{-0.5ex}{$\stackrel{\scriptstyle <}{\scriptstyle \sim}$}}
\newcommand{\gap}%
{\raisebox{-0.5ex}{$\stackrel{\scriptstyle >}{\scriptstyle \sim}$}}

\def\Gr{G}\def\Se{\Sigma}

\def\Ga{\Gr}
\def\Sa{\Se}

\def\Lg{{\cal L}}
\def\Lgh{\makebox[3.5mm]{${\widehat{\makebox[2mm]{$\Lg$}}}$}\vphantom{L}}
\def\Lint{\Lgh^{\mbox{\scriptsize int}}}

\def\A{A}
\def\Gm{\Gamma}
\def\F{F}                             
\def\Ft{\widetilde{F}}                 
\def\Fd{F}                             
\def\Fdt{\widetilde{F}}                
\def\fd{f}

\def\Loop{L}
\def\Ld{\Gamma_{\scr{out}}}
\def\Ldt{\Gamma_{\scr{in}}}
\def\vu{v}

\def\Do{{\cal D}}



%% file: graphs.tex
\def\propagatorR#1#2#3#4{\begin{picture}(1.4,.8)#3
    \if#4r
    \put(1.3,.1){\circle{.2}}\put(1.1,.3){$R$}\else
    \put(.1,.1){\circle{.2}}\put(-0.1,.3){$R$}\fi
    \put(1.3,0.1){\vector(-1,0){.8}}\put(.1,0.1){\line(1,0){.6}}
    \put(.2,-.15){\makebox(0,0){#1}}
    \put(1.2,-.15){\makebox(0,0){#2}}\end{picture}}
\def\propagatorA#1#2#3#4{\begin{picture}(1.4,.6)#3
    \if#4r
    \put(1.3,.1){\circle{.2}}\put(1.1,.3){$R$}\else
    \put(.1,.1){\circle{.2}}\put(-0.1,.3){$R$}\fi
    \put(0.1,0.1){\vector(1,0){.8}}\put(.7,0.1){\line(1,0){.6}}
    \put(.2,-.15){\makebox(0,0){#1}}
    \put(1.2,-.15){\makebox(0,0){#2}}\end{picture}}
\def\self#1#2#3{\begin{picture}(2.2,.8)\thicklines
    \put(2.3,0.1){\vector(-1,0){1.3}}\put(-.1,0.1){\line(1,0){1.3}}
    \put(1.1,0.1){\oval(2,1)}
    \put(0,-0.15){\makebox(0,0){#1}}\put(2.2,-0.15){\makebox(0,0){#2}}
    \if#3R {\put(0.1,0.1){\circle{.2}}}\else
    \if#3A {\put(2.1,0.1){\circle{.2}}}\else
    {\put(1.1,0.1){\makebox(0,0){#3}}}
    \fi\fi\end{picture}}

\linethickness{1.0pt}

\def\gsand4{\begin{picture}(38,0)
\put(2,2){
     \thicklines
     \put(2,0){\circle*{2.00}}
     \put(32,0){\circle*{2.00}}
     \bezier{216}(2,0)(17,-10)(32,0)
     \bezier{216}(2,0)(17,+10)(32,0)
     \bezier{284}(2,0)(17,-25)(32,0)
     \bezier{284}(2,0)(17,+25)(32,0)}
     \end{picture}}
\def\ssand3{\begin{picture}(38,0)
\put(2,2){
     \thicklines
     \put(2,0){\line(1,0){30}}
     \put(2,0){\circle*{4.00}}
     \put(32,0){\circle*{4.00}}
     \bezier{284}(2,0)(17,-25)(32,0)
     \bezier{284}(2,0)(17,+25)(32,0)}
     \end{picture}}
\def\sphisand2phi{\begin{picture}(42,0)
\put(2,2){
     \thicklines
     \put(2,0){
     \put(2,0){\line(1,0){8}}
     \put(24,0){\line(1,0){8}}
     \put(10,0){\makebox(0,0){$\bigotimes$}}
     \put(24,0){\makebox(0,0){$\bigotimes$}}
     \put(2,0){\circle*{4.00}}
     \put(32,0){\circle*{4.00}}
     \bezier{284}(2,0)(16,-25)(32,0)
     \bezier{284}(2,0)(16,+25)(32,0)}}
     \end{picture}}
\def\gphisand3phi{\begin{picture}(53,0)
\put(2,2){
     \thicklines
     \put(3,0){\line(1,0){44}}
     \put(3,0){\makebox(0,0){$\bigotimes$}}
     \put(47,0){\makebox(0,0){$\bigotimes$}}
     \put(8,0){
     \put(2,0){\circle*{2.00}}
     \put(32,0){\circle*{2.00}}
     \bezier{284}(2,0)(16,-25)(32,0)
     \bezier{284}(2,0)(16,+25)(32,0)}}
     \end{picture}}

\def\DSa{\begin{picture}(0,0)\thicklines
\put(0,0){\oval(1.5,1)}
\put(0,0){\makebox(0,0){$-\ii\Sa$}}\end{picture}}
\def\GlnG0Sa{
\begin{picture}(4.3,1.5)
\put(0.5,.1){
\put(.5,0){\DSa}\put(3,0){\DSa}\put(1.75,1){\DSa}
\put(1.75,-1){\makebox(0,0){\dots\dots}}
\put(1,.5){\oval(1,1)[lt]}\put(2.5,.5){\oval(1,1)[rt]}
\put(1,-.5){\oval(1,1)[lb]}\put(2.5,-.5){\oval(1,1)[rb]}}
\end{picture}}

\def\loopb#1{
\unitlength 1.00mm
\thicklines
\begin{picture}(17.50,13.00)
\put(2.00,1.00){\circle*{1.00}}
\put(17.00,1.00){\circle*{1.00}}
\bezier{68}(2.00,1.00)(9.50,5.00)(17.00,1.00)
\bezier{68}(2.00,1.00)(9.50,-3.00)(17.00,1.00)
\bezier{112}(2.00,1.00)(9.50,13.00)(17.00,1.00)
\bezier{112}(2.00,1.00)(9.50,-11.00)(17.00,1.00)
\put(9.93,-1.07){\vector(1,0){0.2}}
\put(9.00,-1.07){\line(1,0){0.93}}
\put(9.93,2.93){\vector(1,0){0.2}}
\put(8.93,2.93){\line(1,0){1.00}}
\put(9.13,6.93){\vector(-1,0){0.2}}
\put(10.07,6.93){\line(-1,0){0.93}}
\put(9.20,-5.00){\vector(-1,0){0.2}}
\put(10.07,-5.07){\line(-1,0){0.87}}
\if#1c \multiput(9.5,-8.)(0,6.5){3}{\line(0,1){5}}\fi
\end{picture}}
\def\loopc#1{ 
\unitlength 1.00mm
\thicklines
\begin{picture}(18.54,9.47)
\put(2.00,-5.00){\circle*{0.93}}
\put(18.00,-5.00){\circle*{1.07}}
\put(10.00,9.00){\circle*{0.94}}
\bezier{72}(2.00,-5.00)(10.00,-1.00)(18.00,-5.00)
\bezier{72}(2.00,-5.00)(10.00,-9.00)(18.00,-5.00)
\bezier{80}(2.00,-5.00)(2.00,6.00)(10.00,9.00)
\bezier{72}(2.00,-5.00)(9.07,-0.80)(10.00,9.00)
\bezier{72}(18.00,-5.00)(11.07,-0.93)(10.00,9.00)
\bezier{76}(18.00,-5.00)(18.13,5.07)(10.00,9.00)
\put(9.93,-7.00){\vector(-1,0){0.2}}
\put(11.00,-7.00){\line(-1,0){1.07}}
\put(3.80,3.27){\vector(1,2){0.2}}
\multiput(3.40,2.47)(0.10,0.20){4}{\line(0,1){0.20}}
\put(16.73,2.33){\vector(2,-3){0.2}}
\multiput(16.20,3.13)(0.11,-0.16){5}{\line(0,-1){0.16}}
\put(10.07,-3.00){\vector(1,0){0.2}}
\put(9.07,-3.00){\line(1,0){1.00}}
\put(12.53,0.60){\vector(-2,3){0.2}}
\multiput(13.00,-0.20)(-0.12,0.20){4}{\line(0,1){0.20}}
\put(7.60,0.67){\vector(-2,-3){0.2}}
\multiput(8.00,1.33)(-0.10,-0.17){4}{\line(0,-1){0.17}}
\if#1c \multiput(8.5,-9.5)(3.5,5.25){3}{\thinlines\line(2,3){2.5}}\fi
\end{picture}}
\def\pia{
\unitlength 1.00mm
\thicklines
\begin{picture}(10.00,8.93)
\bezier{28}(2.00,5.00)(2.13,8.33)(6.00,8.93)
\bezier{28}(2.00,4.87)(2.13,1.53)(6.00,0.93)
\bezier{28}(10.00,5.00)(9.87,8.33)(6.00,8.93)
\bezier{28}(10.00,4.87)(9.87,1.53)(6.00,0.93)
\put(6.00,1.00){\circle*{1.00}}
\put(3.93,1.00){\vector(1,0){0.2}}
\put(2.93,1.00){\line(1,0){1.00}}
\put(6.60,8.80){\vector(1,0){0.2}}
\put(5.60,8.80){\line(1,0){1.00}}
\put(9.83,1.00){\vector(1,0){0.2}}
\put(3.90,1.00){\line(1,0){5.93}}
\end{picture}}
\def\pib{
\unitlength 1.00mm
\thicklines
\begin{picture}(19.78,13.00)
\put(2.00,1.00){\circle*{1.00}}
\put(17.00,1.00){\circle*{1.00}}
\bezier{112}(2.00,1.00)(9.50,13.00)(17.00,1.00)
\put(9.47,1.00){\vector(1,0){0.2}}
\bezier{28}(2.00,1.00)(9.47,1.00)(9.47,1.00)
\put(9.47,6.93){\vector(-1,0){0.2}}
\put(10.07,6.93){\line(-1,0){0.60}}
\bezier{76}(2.00,1.00)(9.60,6.47)(17.00,1.00)
\put(10.00,3.60){\vector(1,0){0.2}}
\put(9.00,3.60){\line(1,0){1.00}}
\bezier{8}(-0.07,1.00)(1.93,1.00)(1.93,1.00)
\put(0.60,1.07){\vector(1,0){0.2}}
\put(-0.00,1.07){\line(1,0){0.60}}
\put(19.78,1.00){\vector(1,0){0.2}}
\put(9.44,1.00){\line(1,0){10.33}}
\end{picture}}
\def\pic{
\unitlength 1.00mm
\thicklines
\begin{picture}(21.78,15.60)
\put(2.07,1.13){\circle*{0.93}}
\put(18.07,1.13){\circle*{1.07}}
\put(10.07,15.13){\circle*{0.94}}
\bezier{80}(2.07,1.13)(2.07,12.13)(10.07,15.13)
\bezier{72}(2.07,1.13)(9.14,5.33)(10.07,15.13)
\bezier{72}(18.07,1.13)(11.14,5.20)(10.07,15.13)
\bezier{76}(18.07,1.13)(18.20,11.20)(10.07,15.13)
\put(3.87,9.40){\vector(1,2){0.2}}
\multiput(3.47,8.60)(0.10,0.20){4}{\line(0,1){0.20}}
\put(16.80,8.46){\vector(2,-3){0.2}}
\multiput(16.27,9.26)(0.11,-0.16){5}{\line(0,-1){0.16}}
\put(12.60,6.73){\vector(-2,3){0.2}}
\multiput(13.07,5.93)(-0.12,0.20){4}{\line(0,1){0.20}}
\put(7.67,6.80){\vector(-2,-3){0.2}}
\multiput(8.07,7.46)(-0.10,-0.17){4}{\line(0,-1){0.17}}
\put(10.07,1.13){\vector(1,0){0.2}}
\bezier{32}(2.07,1.13)(10.07,1.13)(10.07,1.13)
\bezier{8}(-0.00,1.00)(1.00,1.00)(2.00,1.00)
\put(0.73,1.00){\vector(1,0){0.2}}
\put(-0.07,1.00){\line(1,0){0.80}}
\put(21.78,1.11){\vector(1,0){0.2}}
\put(10.22,1.11){\line(1,0){11.56}}
\end{picture}}
\def\wa#1{
\unitlength 0.7mm
\thicklines
\begin{picture}(7,5)\put(2,0){
\put(2.00,1.00){\circle*{1.00}}
\put(2.00,6.00){\makebox(0,0){$ #1 $}}
\bezier{16}(0.54,-0.41)(3.54,2.55)(3.54,2.55)
\bezier{16}(0.54,2.55)(3.50,-0.41)(3.50,-0.41)
\put(1.21,0.22){\vector(1,1){0.2}}
\multiput(0.54,-0.45)(0.11,0.11){6}{\line(0,1){0.11}}
\put(4.61,3.51){\vector(1,1){0.8}}
\multiput(2.32,1.22)(0.11,0.11){20}{\line(0,1){0.11}}
\put(1.25,1.76){\vector(3,-4){0.2}}
\multiput(0.58,2.55)(0.11,-0.13){6}{\line(0,-1){0.13}}
\put(4.54,-1.37){\vector(1,-1){0.8}}
\multiput(2.25,0.76)(0.13,-0.12){18}{\line(1,0){0.13}}
}
\end{picture}}
\def\wb#1{
\unitlength 0.7mm
\thicklines
\begin{picture}(17.51,9.01)\put(4,0){
\put(9.51,-6.49){\circle*{1.00}}
\put(9.51,8.51){\circle*{1.00}}
\put(9.51,-9.49){\makebox(0,0){$ #1 $}}
\put(9.51,11.51){\makebox(0,0){$ #1 $}}
\bezier{68}(9.51,-6.49)(5.51,1.01)(9.51,8.51)
\bezier{68}(9.51,-6.49)(13.51,1.01)(9.51,8.51)
\put(11.58,1.44){\vector(0,1){0.2}}
\put(11.58,0.51){\line(0,1){0.93}}
\put(7.52,0.58){\vector(0,-1){0.2}}
\put(7.52,1.51){\line(0,-1){0.93}}
\bezier{20}(7.00,8.55)(12.04,8.55)(12.04,8.55)
\bezier{20}(7.00,-6.49)(12.04,-6.53)(12.04,-6.53)
\put(8.00,8.51){\vector(1,0){0.2}}
\put(7.00,8.51){\line(1,0){1.00}}
\put(13.44,8.55){\vector(1,0){0.2}}
\put(12.44,8.55){\line(1,0){1.00}}
\put(8.00,-6.58){\vector(1,0){0.2}}
\put(7.00,-6.58){\line(1,0){1.00}}
\put(13.44,-6.58){\vector(1,0){0.2}}
\put(12.44,-6.58){\line(1,0){1.00}}
}
\end{picture}
}